\documentclass[aps,pra,preprint, superscriptaddress,showpacs]{revtex4-1}

\usepackage[pdftex]{graphicx,color}
\usepackage{bm}        
\usepackage{amssymb}   
\usepackage{amsmath}
\usepackage{epsfig} 
\usepackage{threeparttable}
\def\prb#1#2#3{Phys.~Rev.~B~{\bf #1},\ #2\ (#3)}

\def\jcp#1#2#3{J.~Chem.~Phys.~{\bf #1},\ #2\ (#3)}
\def\cpl#1#2#3{Chem.~Phys.~Lett.~{\bf #1},\ #2\ (#3)}

\def\pra#1#2#3{Phys.~Rev.~A~{\bf #1},\ #2\ (#3)}
\def\prl#1#2#3{Phys.~Rev.~Lett.~{\bf #1},\ #2\ (#3)}

\def\rmp#1#2#3{Rev.~Mod.~Phys.~{\bf #1},\ #2\ (#3)}
\def\Eurolett#1#2#3{Europhys.~Lett.~{\bf #1},\ #2\ (#3)}
\def\josab#1#2#3{J.~Opt.~Soc.~Am.~B.~{\bf #1},\ #2\ (#3)}
\def\ao#1#2#3{App.~Opt.~{\bf #1},\ #2\ (#3)}
\def\aamop#1#2#3{Adv.~At.~Mol.~Opt.~Phys.~{\bf#1},\ #2\ (#3)}

\def\sigmarate{\hat{\sigma}^{(2)}}

\def\sigmatwo{\sigma^{(2)}}
\def\klossavg{\bigl<K_\mathrm{loss}\bigr>}
\def\kinavg{\bigl<K_\mathrm{in}\bigr>}
\def\kelavg{\bigl<K_\mathrm{el}\bigr>}
\def\kevapavg{\bigl<K_\mathrm{evap}\bigr>}

\def\SA{S_{\rm A}}
\def\SB{S_{\rm B}}

\def\MSA{M_{S_{\rm A}}}
\def\MSB{M_{S_{\rm B}}}

\def\k1{k_1}
\def\k2{k_2}
\def\q1{q_1}
\def\q2{q_2}

\def\({\left (}
\def\){\right )}
\def\[{\left [}
\def\]{\right ]}

\def\SA{S_{\mathrm{A}}}

\def\MSA{M_{S_{\mathrm{A}}}}

\newcommand{\beq}{\begin{equation}}
\newcommand{\eeq}{\end{equation}}

\newcommand{\threejm}[6]{ \left(\begin{array}{ccc} #1 & #3 & #5\\
                                             #2 & #4 & #6
                               \end{array}
                         \right)}

\def\tA{\tau_\text{A}}
\def\tB{\tau_\text{B}}

\begin{document}
\date{\today}
\title{Collisional properties of cold spin-polarized nitrogen gas: theory, experiment, and prospects as a sympathetic coolant for trapped atoms and molecules} 
\author{T. V. Tscherbul}
\affiliation{Harvard-MIT Center for Ultracold Atoms, Cambridge, Massachusetts 02138}
\affiliation{Institute for Theoretical Atomic, Molecular, and Optical Physics,
Harvard-Smithsonian Center for Astrophysics, Cambridge, Massachusetts 02138}\email[]{tshcherb@cfa.harvard.edu}
\author{J. K{\l}os}
\affiliation{Department of Chemistry and Biochemistry, University of Maryland, College Park, Maryland 20742}
\author{A. Dalgarno}
\affiliation{Harvard-MIT Center for Ultracold Atoms, Cambridge, Massachusetts 02138}
\affiliation{Institute for Theoretical Atomic, Molecular, and Optical Physics,
Harvard-Smithsonian Center for Astrophysics, Cambridge, Massachusetts 02138}
\author{B. Zygelman}
\affiliation{Department of Physics and Astronomy, University of Nevada Las Vegas, Las Vegas, Nevada 89154}
\author{Z.~Pavlovic}
\affiliation{Institute for Theoretical Atomic, Molecular, and Optical Physics,
Harvard-Smithsonian Center for Astrophysics, Cambridge, Massachusetts 02138}
\affiliation{Department of Physics, University of Connecticut, Storrs, Connecticut 06269}
\author{M. T. Hummon}
\affiliation{Harvard-MIT Center for Ultracold Atoms, Cambridge, Massachusetts 02138}
\affiliation{Department of Physics, Harvard University, Cambridge, Massachusetts 02138}
\author{H-I Lu}
\affiliation{Harvard-MIT Center for Ultracold Atoms, Cambridge, Massachusetts 02138}
\affiliation{Department of Physics, Harvard University, Cambridge, Massachusetts 02138}
\author{E. Tsikata}
\affiliation{Harvard-MIT Center for Ultracold Atoms, Cambridge, Massachusetts 02138}
\affiliation{Department of Physics, Harvard University, Cambridge, Massachusetts 02138}
\author{J. M. Doyle}
\affiliation{Harvard-MIT Center for Ultracold Atoms, Cambridge, Massachusetts 02138}
\affiliation{Department of Physics, Harvard University, Cambridge, Massachusetts 02138}

\begin{abstract}

We report a combined experimental and theoretical study of collision-induced dipolar relaxation in a cold spin-polarized gas of atomic nitrogen (N). We use buffer gas cooling to create trapped samples of $^{14}$N and $^{15}$N atoms with densities $(5\pm2) \times 10^{12}$~cm$^{-3}$ and measure their magnetic relaxation rates at milli-Kelvin temperatures. These measurements, together with rigorous quantum scattering calculations based on accurate {\it ab initio} interaction potentials for the $^7\Sigma_u^+$ electronic state of N$_2$ demonstrate that dipolar relaxation in N + N collisions occurs at a slow rate of $\sim$10$^{-13}$ cm$^3$/s over a wide range of temperatures (1 mK to 1 K) and magnetic fields (10 mT to 2 T). The calculated dipolar relaxation rates are insensitive to small variations of the interaction potential and to the magnitude of the spin-exchange interaction, enabling the accurate calibration of the measured N atom density. We find consistency between the calculated and experimentally determined rates. Our results suggest that N atoms are promising candidates for future experiments on sympathetic cooling of molecules. 
\end{abstract}

\maketitle

\clearpage
\newpage

\section{Introduction}

Owing to their unique controllability, cold and ultracold molecular gases hold promise for many research applications, ranging from quantum information processing \cite{Carr,Rabl} and simulation of condensed-matter systems \cite{Carr} to novel constituents of exotic quantum phases \cite{Bloch} and reagents for external field-controlled chemical reactions \cite{Roman}. At present, cold molecular ensembles can be produced by a number of experimental techniques \cite{Carr,JohnCPC,BufferGasCooling,Narevicius,Bas,Schnell,Rempe}, which can be broadly classified as direct and indirect. The direct techniques are based on removing thermal energy from a pre-existing ensemble of molecules via collisional thermalization or time-dependent electromagnetic fields. Among the techniques of this kind are cryogenic buffer-gas cooling \cite{JohnCPC}, velocity filtering \cite{Rempe}, and Stark and Zeeman deceleration \cite{Bas,Schnell,Narevicius}. The indirect cooling methods are based on creating molecules in ultracold atomic gases via photoassociation \cite{Julienne,PA} and sweeping a dc magnetic field across a Feshbach resonance~\cite{Carr}.

While direct cooling methods produce molecules with initial temperatures between $50$ and $200$ mK suitable for cold collision experiments \cite{JohnCPC,Bas,Schnell}, further cooling is required to reach the ultracold regime of interest to applications in condensed-matter physics and quantum information processing \cite{Carr}. This may be accomplished via sympathetic cooling, a technique based on collisional equilibration of thermal energy, which takes place when a gas of molecules is brought into thermal contact with a cold reservoir of atoms. Because sympathetic cooling is driven by elastic collisions, it is a truly general technique, which has found numerous applications in cold atom physics \cite{K,Thywissen,Killian}. Most of the sympathetic cooling experiments performed so far used alkali-metal atoms (typically $^{87}$Rb \cite{K,Thywissen}) because of their easy availability via laser cooling and their attractive collisional properties, which allow for sustainable evaporative cooling down to quantum degeneracy. 


The majority of experiments with cold molecules use permanent magnetic or electrostatic traps to capture molecules in their low-field-seeking Stark or Zeeman states, which are intrinsically unstable and may decay by collisions with background gas atoms. While elastic collisions lead to cooling, inelastic collisions heat the gas, cause trap loss, and shorten the lifetime of trapped molecules. The number of elastic collisions per inelastic collision must be large enough ($>$100) to allow for rapid thermalization while keeping inelastic losses to a minimum. An ideal atomic collision partner ($X$) for sympathetic cooling of magnetically trapped molecules must therefore meet the following acceptability criteria

\begin{enumerate}

\item Be available in copious quantities at low and ultralow temperatures. In particular, atoms with magnetic moments of 1$\mu_B$ or more (where $\mu_B$ is the Bohr magneton) can be loaded in permanent magnetic traps via buffer-gas cooling \cite{BufferGasCooling}, evaporatively cooled to very low temperatures \cite{He}, and co-trapped with molecular species \cite{NH-N}.

\item Have low inelastic $X$-$X$ collision rates, so that sufficient density of $X$ can be maintained in the trap at all temperatures to allow for efficient molecule thermalization; 

\item Have low inelastic collision rates with the diatomic molecules of interest, so that elastic atom-molecule collisions which drive thermalization occur more frequently than inelastic collisions.

\end{enumerate}
 

Previous theoretical work has addressed the optimal choice of atomic collision partners for sympathetic cooling of molecular species \cite{Lara,Gianturco,HutsonPRA}. The alkali-metal atoms, which satisfy the requirements (1) and (2), have so far received most attention and {\it ab initio} calculations of interaction energies and low-temperature collision properties have been reported for the alkali-metal atoms Rb and Cs interacting with OH \cite{Lara}, NH \cite{Gianturco}, and ND$_3$ \cite{HutsonPRA}. These studies have shown that the interactions between the alkali-metal atoms and diatomic molecules are strongly anisotropic, giving rise to large inelastic collision rates at low temperatures, thereby limiting the efficiency of sympathetic cooling \cite{HutsonPRA,Gianturco}. A recent theoretical study has shown that sympathetic cooling of OH($^2\Pi$) molecules in low-field-seeking Zeeman states of $e$ symmetry may be facilitated by superimposed  electric and magnetic fields \cite{FD}. We note that certain trapping techniques employing ac electric \cite{ACtrap}, optical dipole \cite{BarkerNJP} or microwave \cite{DeMille} fields, allow for trapping ground-state molecules, thereby eliminating the possibility of collisional relaxation. At their present stage of development, however, these techniques are less advanced than magnetic or electrostatic trapping \cite{Carr}. 


Recently, Wallis and Hutson theoretically explored the possibility of using the alkaline-earth atoms to sympathetically cool paramagnetic molecules \cite{MgNH}. Their {\it ab initio} calculations on Mg~+~NH collisions in the presence of an external magnetic field demonstrated that inelastic collisions are suppressed at low magnetic field strengths, indicating that sympathetic cooling of NH molecules by collisions with laser-cooled Mg atoms might be possible \cite{MgNH}. However, the detrimental inelastic collisions become very efficient at large magnetic fields, which requires precooling of molecular gas to very low temperatures (1 mK) before the sympathetic cooling can begin. In addition, the alkaline-earth atoms in their electronic ground states are not paramagnetic, which makes it challenging to produce large numbers of Mg atoms required for collisional thermalization (criterion 1). The same limitation applies to recent proposals for using laser-cooled rare gas atoms to sympathetically cool large molecules like benzene \cite{BarkerNJP}.

We have recently suggested that molecular species bearing magnetic moments can be sympathetically cooled by collisions with spin-polarized nitrogen (N) atoms in a permanent magnetic trap \cite{NH-N,MattThesis}. Due to their large magnetic moments of 3$\mu_B$, the N atoms can be efficiently confined in a magnetic trap using buffer-gas cooling \cite{BufferGasCooling}. Previous experimental work \cite{NH-N} has demonstrated that samples of N atoms with densities $\sim$10$^{10}$ - 10$^{11}$ cm$^{-3}$ can be routinely produced and held in a magnetic trap for as long as 10 s, allowing for co-trapping of molecular species such as NH \cite{NH-N}. However, the cross sections for inelastic relaxation in N + N collisions were not measured in these preliminary experiments, leaving the question open of whether N + N collisional thermalization would prevail over two-body inelastic losses (criterion 2). While it is well-known that the two-body losses in doubly spin-polarized atomic gases are induced by the magnetic dipole interaction \cite{Pfau}, the time scale for these processes in N + N collisions is unknown. In addition, the density of trapped N atoms could not be accurately determined due to the difficulties encountered in N atom detection.


Here we present a combined experimental and theoretical study of low-temperature collisions in a cold spin-polarized gas of atomic nitrogen. We use buffer-gas cooling to trap large numbers of $^{14}$N and $^{15}$N atoms and study their collision-induced dipolar relaxation at milli-Kelvin temperatures. To interpret the experimental observations, we perform accurate {\it ab initio} calculations of the interaction potential between two spin-polarized N atoms and rigorous quantum scattering calculations of trap loss dynamics. By analyzing various sources of uncertainty in our theoretical results, we infer the upper and lower bounds to the calculated relaxation rates, which allows us to calibrate the N atom density based on the experimental measurements of trap lifetimes.

The paper is organized as follows. In Sec. II, we describe our experimental apparatus and present measurements of collision-induced trap loss rates for both $^{14}$N and $^{15}$N isotopes of atomic nitrogen. Sections IIIA and IIIB present {\it ab initio} calculations of the interaction potentials for N$_2$ and give a brief outline of quantum scattering calculations on N + N collisions. Sections IIC and IIIC compare our theoretical results with experimental data. Section IV gives a brief summary of our results and outlines possible future research directions.



\section{Experiment}

\subsection{Apparatus}
\begin{figure}
\includegraphics[width = .97\textwidth, trim = 0 0 0 -40]{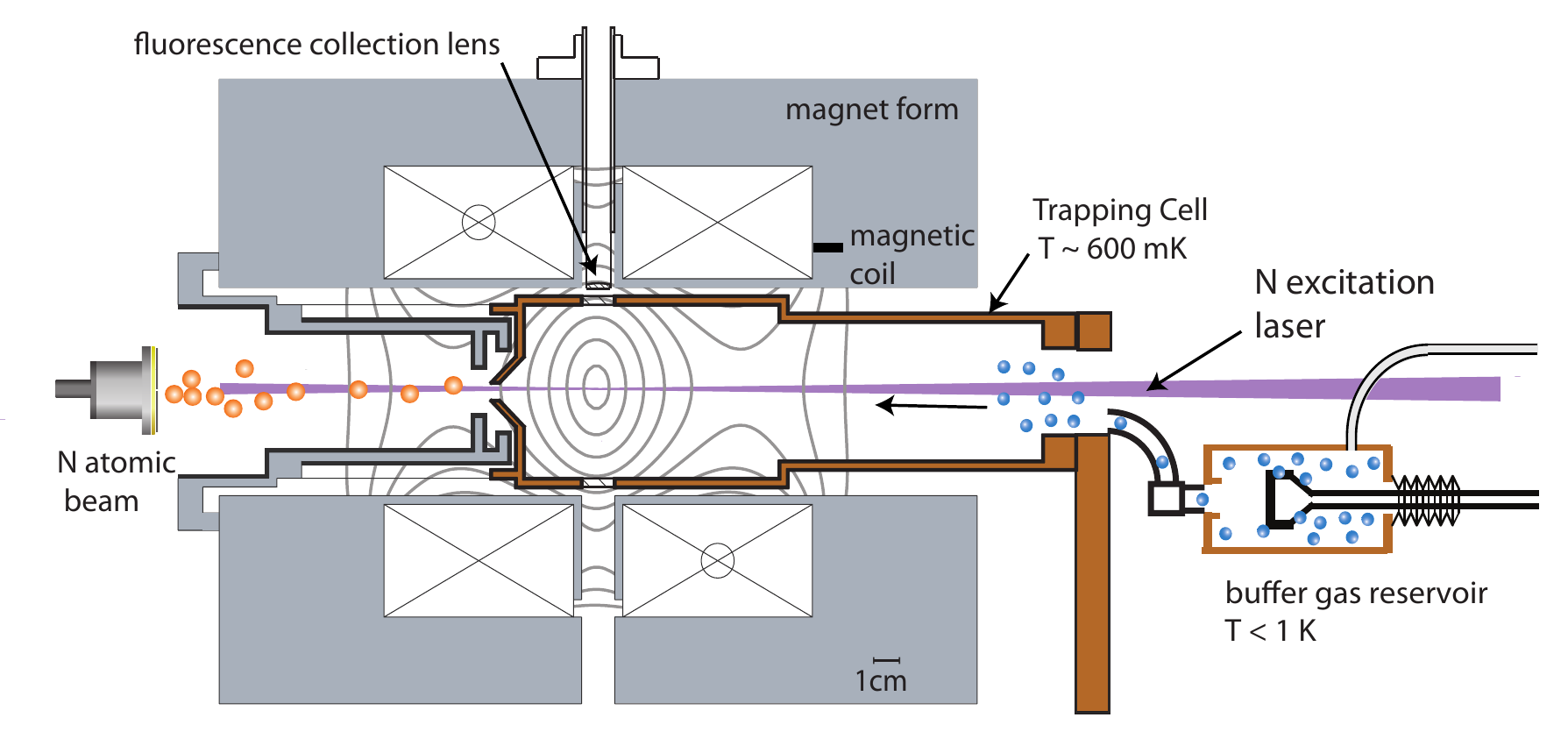}
	\renewcommand{\figurename}{Fig.}

\caption{Diagram of trapping apparatus.}
\label{fig:apparatus}
\end{figure}

The experimental apparatus is similar to those described in Refs.~\cite{Tsikata10,NH-N}.  A diagram of the trapping region is shown in Fig.~\ref{fig:apparatus}.  The trapping region is centered about a pair of super-conducting magnetic solenoids that produce a spherical-quadrupole magnetic trap with depth of 3.8~T.  For atomic nitrogen, with a magnetic moment of $3\mu_\mathrm{B}$, the corresponding trap depth is about 7.6~K.  The magnetic field contours are shown in gray in Fig.~\ref{fig:apparatus}.  In the bore of the super-conducting solenoids resides a cylindrical copper buffer gas cell, maintained at a temperature of about 600~mK by thermal anchoring to a $^3$He refrigerator.  The buffer gas cell has an aperture at both ends of the trapping region.  At one end, a 1~cm diameter aperture allows the atomic nitrogen to enter the trapping region from a room temperature atomic beam.  At the opposite end, a 3.80~cm diameter aperture allows the buffer gas to be rapidly introduced into and subsequently removed from the trapping region.  The atomic nitrogen beam is generated using a DC glow discharge source with N$_2$ as the process gas, operating at a stagnation pressure of 100~torr .  The atomic source is turned on for approximately 40~ms to load atoms into the trapping cell.  Simultaneous with the introduction of atoms into the trapping region, $^3$He buffer gas is introduced into the cell by pulsing open the cryogenic buffer gas reservoir \cite{Tsikata10}.  The density of the buffer gas during loading of the atoms is on the order of $10^{15}$~cm$^{-3}$.  The loading pulse of buffer gas  then exits the buffer gas cell via the 3.80~cm aperture on a time scale of 50~ms.  Following loading of N into the magnetic trap, the final background buffer gas density in the trapping region is approximately $10^{12}$~cm$^{-3}$.  The density is set by the rate of helium desorbing from the buffer gas cell walls \cite{Harris:04}.  Nitrogen-helium collisions, occuring at a rate of about 100~Hz due to residual buffer gas, pin the trapped nitrogen temperature to the temperature of the cell walls.  This allows the trapped nitrogen temperature to be monitored using a ruthenium oxide thermometer mounted to the cell wall.  A resistive heater mounted to the cell wall allows the nitrogen temperature to be adjusted. The nitrogen trap loss lifetime due to these N-He elastic collisions is on the order of 100~s since the temperature of the He buffer gas is more than a factor of 10 lower than the depth of the magnetic trap. This long trap lifetime makes it possible to study N-N collisions that lead to trap loss on time scales of 10~s.  

\subsection{Atomic nitrogen detection}
To detect the trapped atomic nitrogen we use two-photon absorption laser induced fluorescence (TALIF) \cite{adams_TALIF}.  We excite atomic nitrogen in the ground $(2p^3)^4$S$_{3/2}$ state by absorbing two photons at 206.7 nm to the excited $(3p)^4$S$_{3/2}$ state at 96750 cm$^{-1}$.  The excited $(3p)^4$S$_{3/2}$ state has a lifetime of 26~ns \cite{bengtsson_n_lifetime} and decays to the $(3s)^4$P states, emitting light near 745~nm.  A 1~m focal length lens placed outside the vacuum chamber is used to focus the excitation laser onto the trapped sample.   The fluorescence is collected using a lens mounted at the midplane of the magnet and sent to a photomultiplier tube for detection.

Estimation of the trapped atomic nitrogen density from the TALIF signal is difficult.  Both the fluorescence collection efficiency and nitrogen excitation probability are required to convert the TALIF signal to an absolute nitrogen density.  From geometric considerations and fluorescence measurements using trapped NH, we estimate our fluorescence collection efficiency to be $10^{-4}$ \cite{MattThesis}.  This value for fluorescence collection effieciency we estimate to be accurate to within a factor of 3.  To calculate the nitrogen excitation probability, one needs precise knowledge of the spatial, temporal, and spectral properties of the excitation laser.  In the low intensity limit, where depletion of the ground state and photo-ionization are negligible, one can show that the total number of fluorescence photons produced from the sample is \cite{Bamford_O}:
\begin{align}
N_\mathrm{photon} & = \sigmarate \frac{E^2}{(h\nu)^2}\int n_\mathrm{gr}(r) S^2(r) dV  \int_{-\infty}^{\infty} F^2(t)dt
	\label{eq:photon_number}
\end{align}
where $N_\mathrm{photon}$ is the total number of fluorescence photons produced per pulse, $\sigmarate$ is the effective two-photon cross section , $n_\mathrm{gr}(r)$ is the nitrogen ground state density distribution, $E$ is the laser pulse energy, $h\nu$ is the excitation laser photon energy, and $S(r)$ and $F(t)$ are the normalized spatial and temporal profiles of the laser beam ($\int S(r) dA = 1$, and $\int F(t)dt = 1$).  Here we have assumed the spatial and temporal variations are independent.  We can express the value of the spatial integral in terms of an effective $1/e^2$ beam waist $w_0$, where $\int S(r)^{2} dA = \pi^{-1} w_0^{-2}$.  Similarly, the temporal integral can be expressed in terms of an effective pulse duration, $\tau_{\mathrm{ex}}$, where $\int F(t)^2 dt = \sqrt{2\ln(2)\pi^{-1}} (\tau_\mathrm{ex})^{-1}$. The resonant effective two-photon cross section can be expressed as \cite{BAMFORD:1988ff}:
\begin{align}
\sigmarate = \sigmatwo g(\delta = 0)G^{(2)}(t =0)
\end{align}
where $g(\delta =0)$ is the resonant line shape factor, $G^{(2)}(t = 0)$ is the second-order intensity-correlation function of the excitation laser, $\delta$ is the detuning from the atomic resonance, $\delta = 2\omega_\mathrm{laser} - \omega_0$, and the line shape factor is normalized such that $\int g(\delta) d\delta = 1$.

The spatial profile of the laser at the position of the atoms is measured using a CCD camera. The value of $\int S(r)^2dA$ is measured to be $ \pi^{-1} (120\mu $m)$^{-2}$.  The term $\int n_\mathrm{gr}(r) S^2(r) dV$ in Eq.~(\ref{eq:photon_number}) can be evaluated in the following manner.  For the trap geometry in our experiment, the trapped nitrogen density does not vary significantly over the spatial profile of the laser.  In the direction of propagation of the laser, the nitrogen atoms are confined to a effective length, $l_\mathrm{eff} = 2$~mm.  The term can then be evaluated $\int n_\mathrm{gr}(r) S^2(r) dV = n_0 l_\mathrm{eff} \int S(r)^2dA$, where $n_0$ is the nitrogen density at the center of the trap.   

To monitor the temporal profile of the laser we pick off a portion of the laser beam and direct it onto a ceramic beam dump.  We then use a fast photodiode to monitor the light scattered from the beam dump. The value of  $\int F(t)^2 dt$ is measured to be $ \sqrt{2\ln(2)\pi^{-1}} (9.5 $ns$)^{-1}$. The photodiode is also used to monitor the laser pulse energy, $E$, by calibrating its signal using a pyroelectric energy meter.  

\begin{figure}
\includegraphics[width = .8\textwidth, trim = 0 0 0 -40]{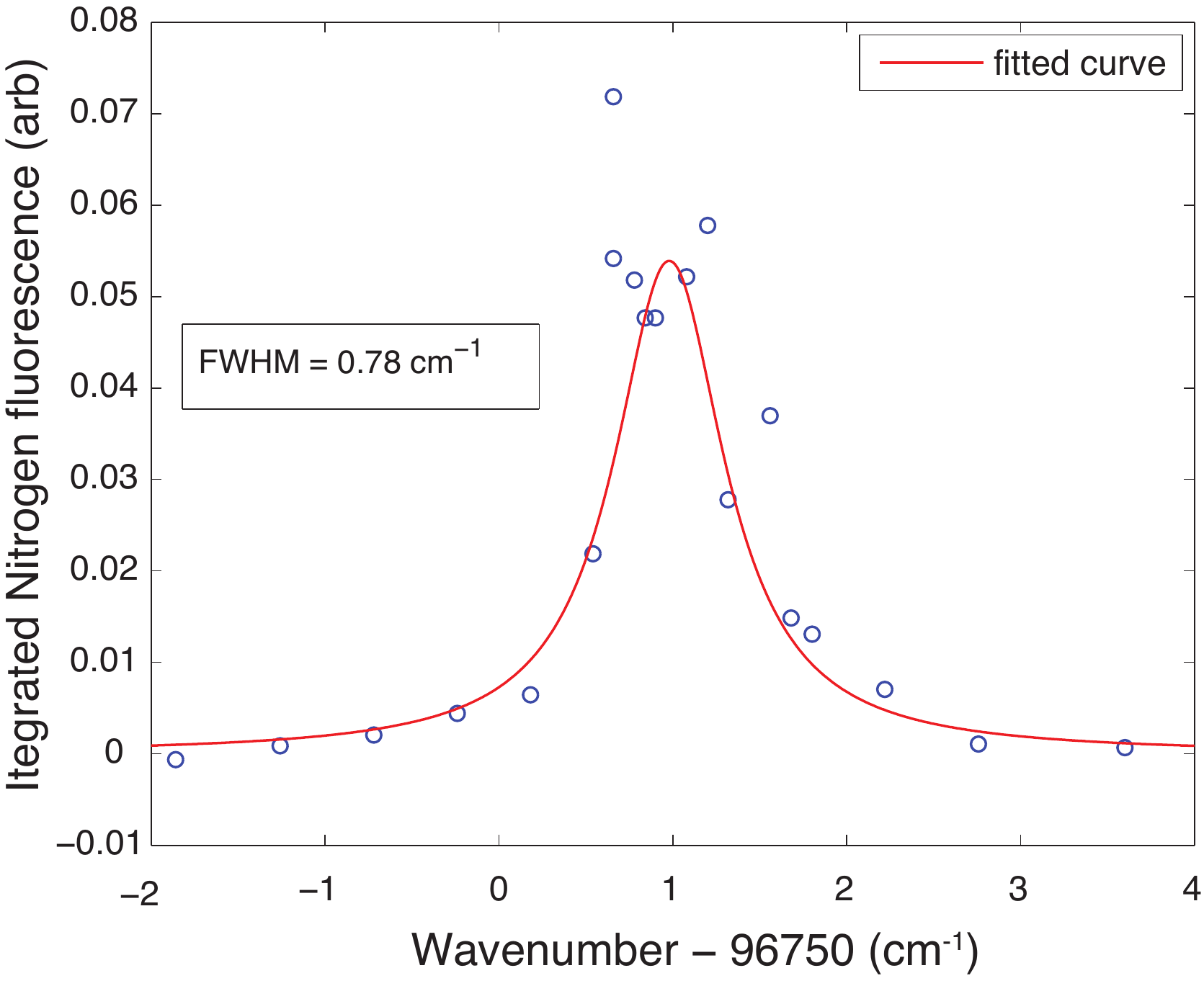}
	\renewcommand{\figurename}{Fig.}

\caption{Trapped nitrogen spectrum, fitted to a Lorentzian profile.}
\label{fig:N_spectrum}
\end{figure}

The spectral profile of the pulsed laser is more difficult to characterize.  The commercial Sirah pulsed dye laser uses a Littman-Metcalf configuration for the resonator cavity  \cite{littman1978spectrally} to produce laser light at 620~nm.  The resonator cavity has a linewidth of about 1.5 GHz and longitudinal mode spacing of about 600 MHz.  For each laser pulse, several different longitudinal modes may lase.  This is observed in our setup using a Fizeau interferometer  \cite{KAJAVA:1992bs,WESTLING:1984ta} as a spectrum analyzer.  The 620~nm light is subsequently frequency doubled in a KDP crystal, and the doubled light mixed with the fundamental in a BBO crystal to produce light at 207~nm.  The shot to shot spectral variation of 1.5 GHz at 620~nm leads to a variation of 9~GHz (0.3 cm$^{-1}$) at the resonant atomic frequency.  For comparison, the expected Doppler broadening (full-width half-max) of the atomic transition is expected to be 300 MHz at 600~mK with Zeeman broadening of less than 100 MHz \cite{MooreV1,HIRSCH:1977eq}.  As a result, the value of $g(0)$ will be determined by the spectral properties of the laser.  An upper limit on $g(0)$ can be determined by directly measuring the observed nitrogen signal linewidth.  Figure~\ref{fig:N_spectrum} shows a trapped nitrogen spectrum taken at a cell temperature of 600~mK.  To acquire these spectra, we monitor the wavelength of the excitation laser using a wavemeter with resolution 0.01~cm$^{-1}$ at 620~nm, corresponding to a resolution of 0.06 cm$^{-1}$ at the atomic transition frequency. Each data point represents the average signal of three consecutive laser excitation pulses.  These consecutive laser pulses have a corresponding frequency jitter on the order of 0.3 cm$^{-1}$ at the atomic transition frequency.  The measured linewidth of 0.76~cm$^{-1}$ (23 GHz) is a result of a combination of the actual spectral linewidth of the laser and the limited resolution of our measurement technique.  For calculations in this paper, we take $g(0) = (2/\pi) ( 2\pi\times10~$GHz$)^{-1}$. This value of $g(0)$ should be good to a factor of 2. 


Finally, due to the complicated spectral mode structure of the light at 620~nm, it is not clear what the value of $G^{(2)}(t = 0)$ is for the light produced at 207~nm.  In Ref.~\cite{BAMFORD:1988ff}, Bamford and coworkers analyze values of $G^{(2)}(t = 0)$ for various pulsed laser systems. They find $G^{(2)}(0)$ ranges between 1.5 and 3.0 for typical multimode pulsed laser systems, though our setup is not directly measured in the reference.  Taking a value of $G^{(2)}(0) = 2$ will be within $30\%$ of the actual value.	

The parameters to calibrate the TALIF signal are summarized in Table~I.  Due to the large uncertainties in the spectral properties of the excitation light and the fluorescence collection efficiency, the nitrogen densities calculated using the values listed in Table~I should be accurate only to within an order of magnitude. Also omitted from this analysis is the role of laser polarization and atom orientation, which would likely introduce corrections of order unity to the calculation.

\subsection{Experimental results}
\label{sec:Experimental_results}

Here we present our observations of trapping of atomic nitrogen and discuss the nature of the observed trap loss.  In particular, we are interested in the ratio of the elastic N + N collision rate to the inelastic N + N collision rate, $\gamma$.  Measurements of both the elastic and inelastic collision rates are desirable, though to measure each independently, one needs to have an absolute calibration of the atomic density.  Since our estimates of atomic N density from the TALIF signal are only good to about an order of magnitude, we lack the information we need to make precise measurements of the elastic and inelastic N + N collision rates.

However, it is possible to directly measure $\gamma$ without precise knowledge of our atomic nitrogen density by investigation of the dynamics of the magnetic trap loss.  Qualitatively, for very deep traps, evaporation of the sample is suppressed, and loss is driven by inelastic N + N collisions.  For lower trap depths, evaporative loss due to elastic N + N collisions can dominate trap loss.  By measuring trap loss due to N + N collisions over a range of trap depths, it is possible to directly extract $\gamma$.   A discussion of our magnetic trap dynamics follows.

The expression for loss of  atomic N in our magnetic trap has the form (see Appendix~A)
\begin{align}
\dot{n}_0 =-  \frac{1}{7.6} \bigl< K_\mathrm{loss}\bigr> n_0^2 - \frac{1}{\tau_\mathrm{He}} n_0 \label{eq:combined_loss}
\end{align}
where $n_0$ is the peak trap density, $\klossavg$ is the trap average 2-body loss rate coefficient, $1/7.6$ is a factor for our trap geometry, and $\tau_\mathrm{He}$ is the $1/e$ lifetime associated with loss due to atom-helium collisions. $\klossavg$ includes trap loss both from atom evaporation, $\kevapavg$, and loss from atom-atom inelastic collisions, $\kinavg$.  The evaporative portion of trap loss can be expressed as $\kevapavg = f(\eta)\kelavg$, where $f(\eta)$ is the fraction of elastic collisions that lead to atom loss at trap depth $\eta =T_\mathrm{trap}/T_\mathrm{atom}$, where $T_\mathrm{trap}$ is the trap depth expressed in units of temperature, and $T_\mathrm{atom}$ is the temperature of the trapped atomic sample. For our trap geometry, Monte-Carlo simulations of trap dynamics yield $f(\eta) = 1.9(\eta - 3)\exp(-\eta)$, which agrees well with analytic expressions for $f(\eta)$ \cite{Ketterle:1996bv}. The relationship between $\kinavg$ and the dipolar relaxation rates derived from the quantum scattering calculations in Sec. \ref{sec:Theory} is given in Appendix A. Combining these expressions, we have:
\begin{align}
\klossavg &= f(\eta)\kelavg + \kinavg\\
      		& = \kelavg(f(\eta) + 1/\gamma) \label{eq:gamma_fit}
\end{align}
Equation (\ref{eq:gamma_fit}) provides an expression for extracting $\gamma$ without precise knowledge of the absolute atomic nitrogen density.

\begin{figure}
\includegraphics[width = .7\textwidth, trim = 0 0 0 -40]{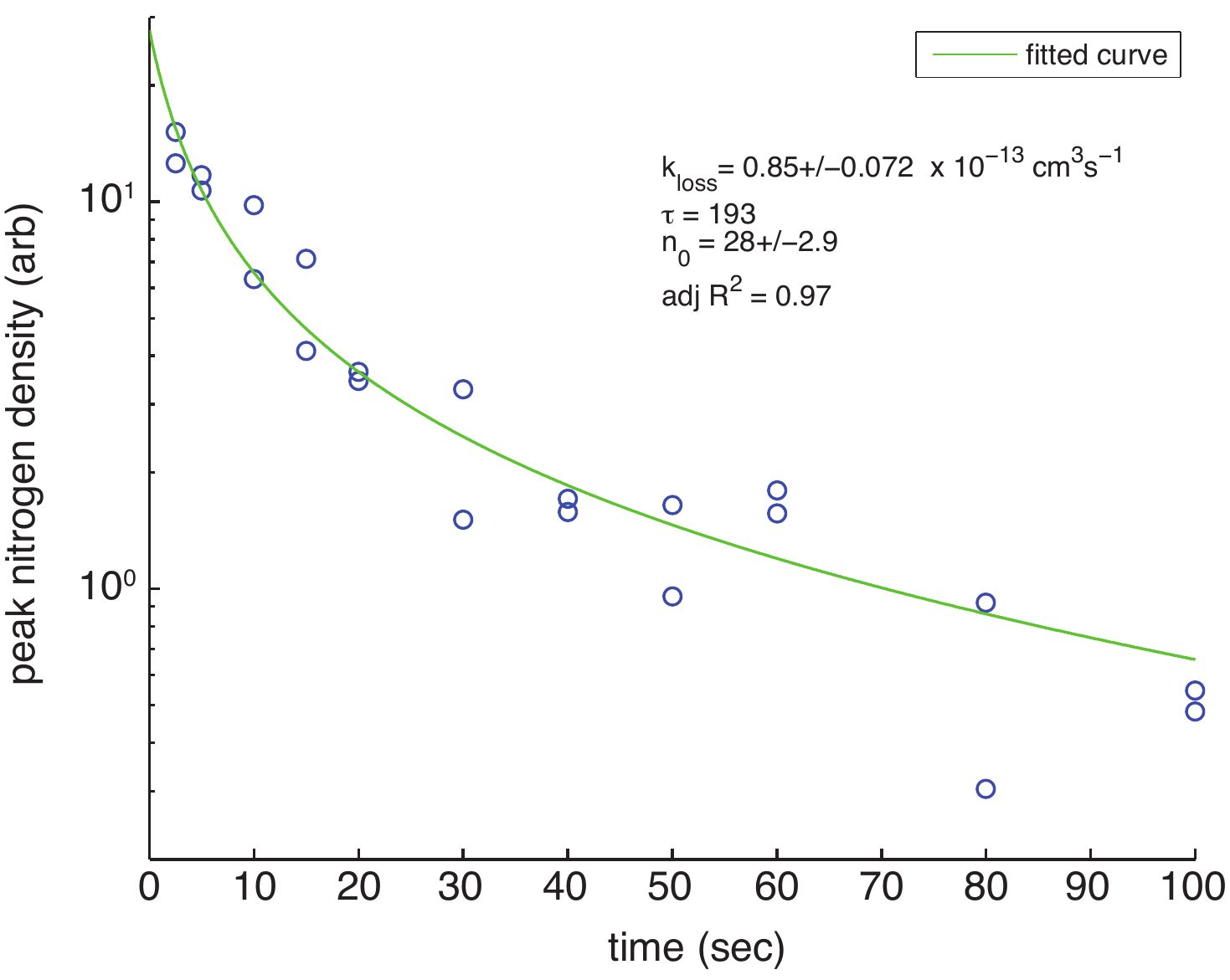}
	\renewcommand{\figurename}{Fig.}

\caption{Time decay of trapped nitrogen.  Initial trap loading occurs at $t=0$.  The trap loss is fitted to the solution of Eq.~\ref{eq:combined_loss} to determine the trap loss parameters. }
\label{fig:N_time_decay}
\end{figure}

Figure \ref{fig:N_time_decay} shows a typical nitrogen trap decay.  For each data point in Fig. \ref{fig:N_time_decay} we load N atoms into the trap at $ t = 0$ and detect the remaining N atoms after waiting a period of time between 2~s and 100~s.  Attempts to continually detect trapped N atoms during a single trap loading result in rapid N trap loss, most likely due to optical pumping.  The shot to shot variation in N signal is likely due to the variation of the spectral properties of the excitation laser.  We fit our nitrogen time decay data to the solution of Eq.~(\ref{eq:combined_loss}) to arrive at values for $\bigl< K_\mathrm{loss}\bigr>$. 



\begin{figure}
\includegraphics[width = .7\textwidth, trim = 0 0 0 -40]{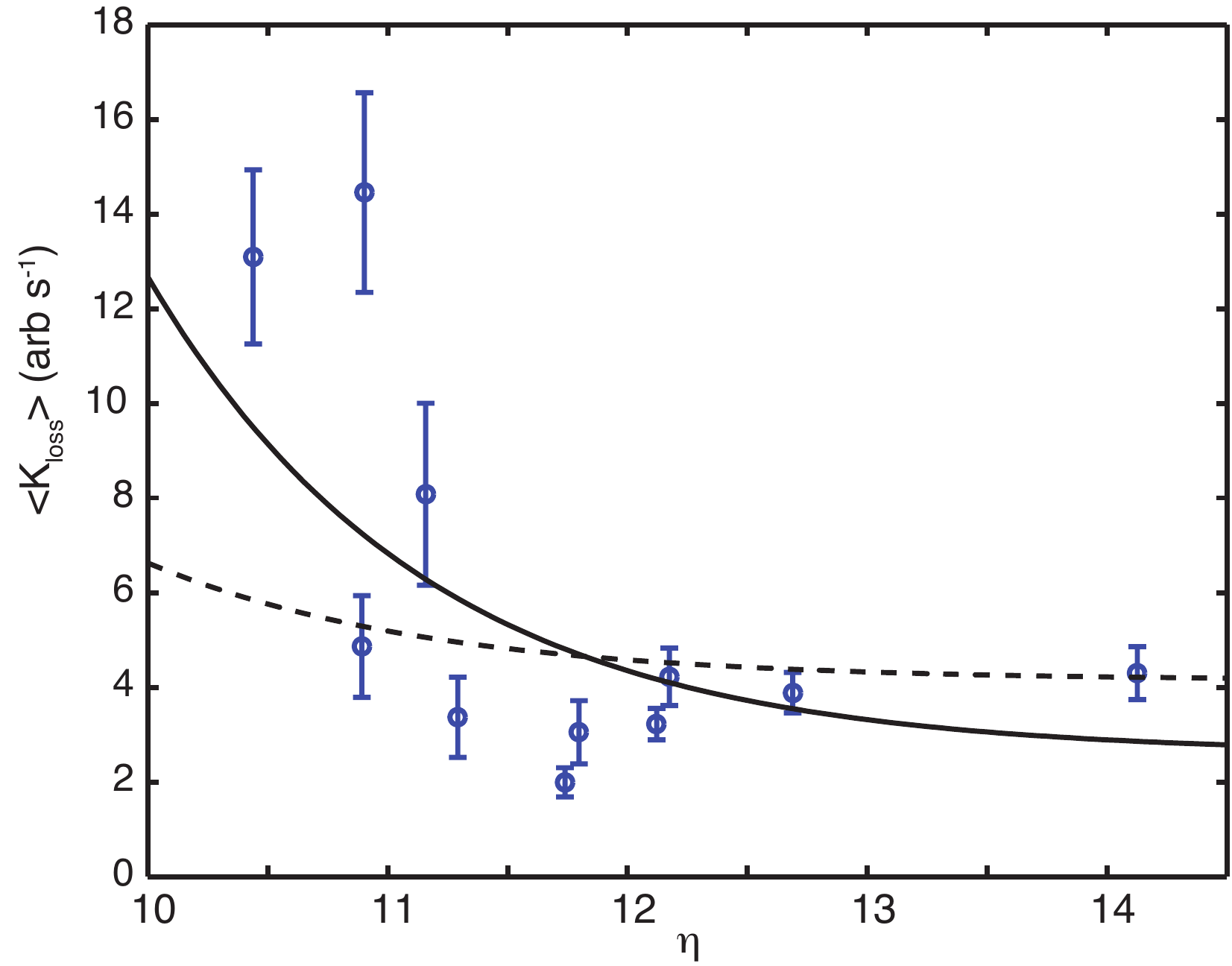}
	\renewcommand{\figurename}{Fig.}

\caption{$\klossavg$ vs $\eta$.  The solid line is a fit of the data to Eq. \ref{eq:gamma_fit}.  The dashed line is a fit of the data to \ref{eq:gamma_fit} with $\gamma$ set to the value calculated in Sec. III, $\gamma_\mathrm{theory} = 1000$.}
\label{fig:n_loss_vs_eta}
\end{figure}
\begin{figure}[t]
	\centering
     \includegraphics[width=0.75\textwidth, trim = 0 0 0 0]{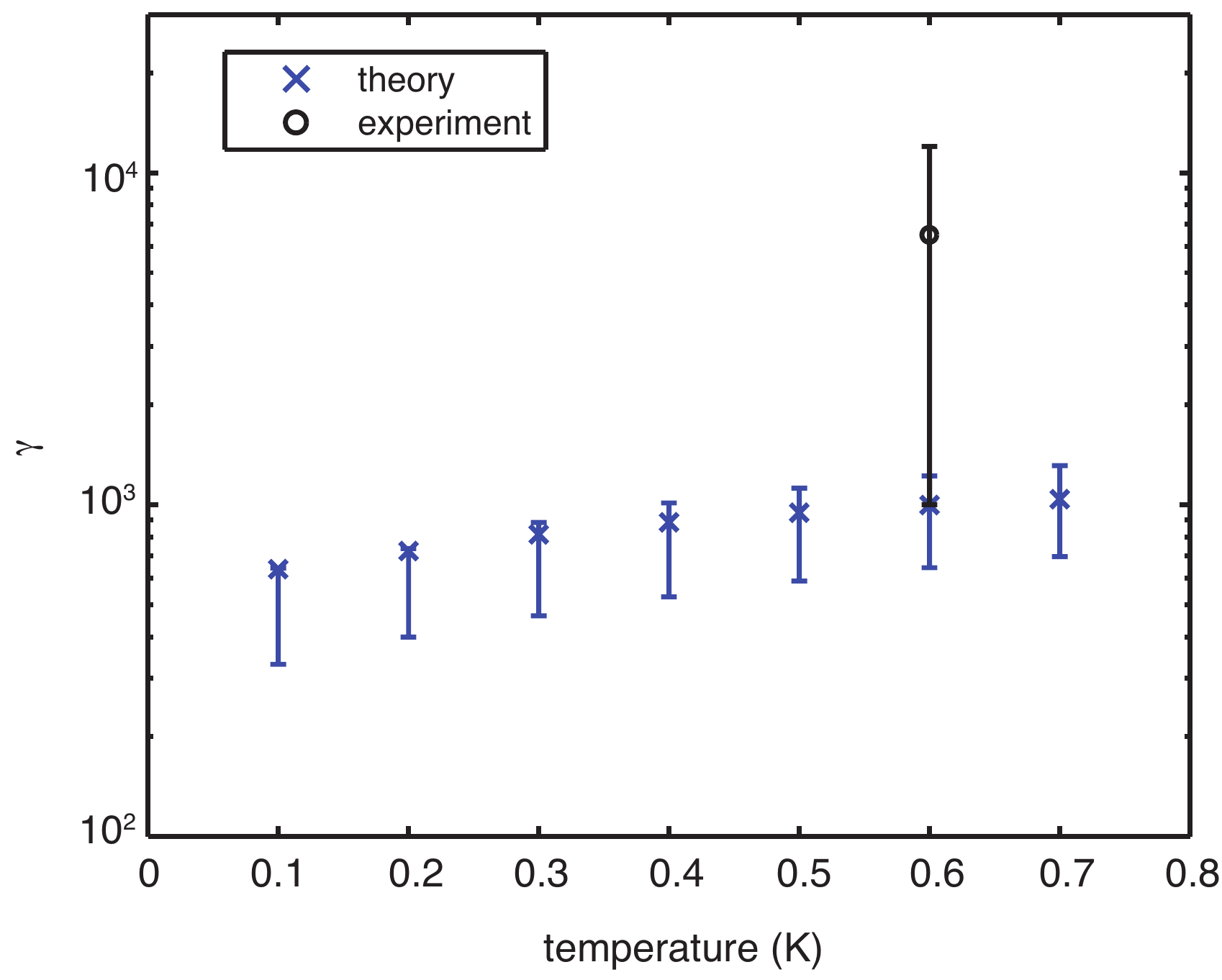}
	\renewcommand{\figurename}{Fig.}
	\caption{The calculated (crosses) and measured (circles) ratios $\gamma = \langle K_\text{el}\rangle/ \langle K_\text{in}\rangle$ for $^{14}$N. The theoretical ratios and error bars are calculated based on the data shown in Figs. 15 and 16 as described in Sec.~IIIC. }
	\label{fig:gamma}
\end{figure}

Values for  $\bigl< K_\mathrm{loss}\bigr>$ are measured for values of $\eta$ between 10 and 14, which correspond to $T_\mathrm{atom}$ between 550~mK and 650~mK with magnetic trap depths between 3.3~T and 3.9~T.  A plot of $\klossavg$ versus $\eta$ is shown in Fig. \ref{fig:n_loss_vs_eta}. The solid line in Fig.~\ref{fig:n_loss_vs_eta} is the fit to Eq. (\ref{eq:gamma_fit}) with $\kelavg$ and $\gamma$ allowed to vary as fit parameters.  This yields a value of $\gamma_\mathrm{exp} = (6.5\pm 5.5) \times 10^3$. Figure \ref{fig:gamma} shows the results of quantum scattering calculations described in the following sections, which give $\gamma_\mathrm{theory} = (1.0\pm0.3)\times10^3$ at $T=600$~mK. The dashed line in Fig.~\ref{fig:n_loss_vs_eta} shows a fit of Eq. (\ref{eq:gamma_fit}) with the value of $\gamma$ fixed to $\gamma_\mathrm{theory} = 1000$, where the only free fitting parameter is $\kelavg$. The values of  $\gamma_\mathrm{exp}$ and $\gamma_\mathrm{theory}$ are consistent, and the value of $\gamma>1000$ is favorable for evaporative cooling of atomic N in a magnetic trap.
 
From the data in Fig.~\ref{fig:n_loss_vs_eta}, one can calibrate the trapped atomic nitrogen density by fitting the observed loss rates to the calculated loss rates.  What is needed is simply a scaling factor, $n_0 = c_\mathrm{N}n_\mathrm{obs}$, to convert the observed nitrogen signal, $n_\mathrm{obs}$, to an actual nitrogen density, $n_0$.  Substitution of the scaling factor relation into Eq.~\ref{eq:combined_loss} results in the expression: 
\begin{align}
\dot{n}_\mathrm{obs} =-  \frac{1}{7.6} \bigl< K_\mathrm{loss}\bigr> c_\mathrm{N}n_\mathrm{obs}^2 - \frac{1}{\tau_\mathrm{He}} n_\mathrm{obs} \label{eq:combined_loss_obs}
\end{align}
By setting the value of $\klossavg$ to the theoretical value calculated as described in Sec. III and allowing $c_\mathrm{N}$ to vary as a fit parameter, one can then fit the solution of Eq.~\ref{eq:combined_loss_obs} to the observed nitrogen trap decay to arrive at a value for the scaling factor $c_\mathrm{N}$.
When using this calibration technique we need to consider the systematic uncertainties associated with arriving at theoretical value of $\klossavg$.  In particular, for $\eta > 12$, $\kinavg$ accounts for $60\%$ or $90\%$ of the total $\klossavg$ for values of $\gamma = 6500$ or 1000, respectively. Although there may be large uncertainty in the actual value of $\gamma$, for large $\eta$, the uncertainty in the systematic correction of $\kinavg$  to arrive at a total $\klossavg$ is only about $30\%$.  From this calibration method we estimate we typically trap atomic nitrogen at initial peak densities of more than $(5\pm2)\times 10^{11}$ cm$^{-3}$, corresponding to more than $(3\pm1) \times 10^{11}$ trapped nitrogen atoms.  The error in these nitrogen density measurements is dominated by the systematic uncertainties associated with the model of trap loss dynamics and the quantum scattering calculations of $\klossavg$, but also includes the statistical uncertainties associated with the experimental measurement of $\klossavg$ (or equivalently $c_\mathrm{N}$) .  The atomic nitrogen density calculated using this method agrees with our nitrogen density estimates from the TALIF signal to within a factor of 5, consistent with our expected uncertainty in the TALIF signal of an order of magnitude. This technique for measurement of the trap nitrogen density is valuable due to technical difficulties associated with a direct spectroscopic measurements of trapped atomic~N. 



We also observe trapping of the bosonic isotope $^{15}$N by using isotopically enriched ($>98\%+$) $^{15}$N$_{2}$ as the process gas.  No differences in trap loss were observed between $^{15}$N and $^{14}$N at a trap temperature of 600~mK.

\section{Theory}
\label{sec:Theory}

\subsection{Ab initio calculations of interaction potentials}

To evaluate the potential energy curve (PEC) for the $^7\Sigma_u^+$ electronic state of N$_2$, we use two different {\it ab initio} approaches. The first approach is based on the coupled cluster method including perturbative triple excitations (CCSD(T)) and extrapolation to the complete basis set (CBS) limit. The second approach uses the coupled cluster method with  full iterative triple excitations (CCSDT) and a fixed basis set with additional bond functions. The interaction energies in both methods are calculated within the supermolecular approach, where dimer and monomer energies are calculated with dimer centered basis sets. We applied Boys and Bernardi counterpoise procedure to correct for the basis set superposition error.

 In the first approach, the PEC was calculated using a single-reference restricted Hartree-Fock (RHF) wave function as a starting  point, followed by a spin-unrestricted coupled cluster treatment~\cite{knowles:93} with single, double and non-iterative triple excitations (UCCSD(T)) as implemented in the MOLPRO suite of programs~\cite{molpro}. The use of the single-reference approach is justified because the high-spin electronic state  $^7\Sigma_u^+$ can be well described by a single determinant. For the purpose of extrapolation to the CBS limit, we used a series of augmented, correlation-consistent triple-, quadruple-, quintuple- and sextuple-zeta basis sets of Dunning {\em et al.}~\cite{dunning:89,kendall:92} denoted as AVTZ, AVQZ, AV5Z and AV6Z, respectively. The $1s$ orbitals of N were frozen in these calculations. The interaction energies were calculated at 50 internuclear separations from $R=2.5$ $a_0$ to $R=50$ $a_0$ and fit to analytic functions of $R$ using the reproducing kernel Hilbert space (RKHS) method~\cite{ho:96, EPAPS}. The radial kernel was composed of the short range part and the asymptotic long-range part proportional to $R^{-6}$. The smoothness of the one-dimensional kernel parameter was set to 2 to allow for a smooth extrapolation of {\it ab initio} data points to the asymptotic region ($R>50a_0$).

\begin{figure}[t]
	\centering
     \includegraphics[width=0.7\textwidth, trim = 0 0 0 -40]{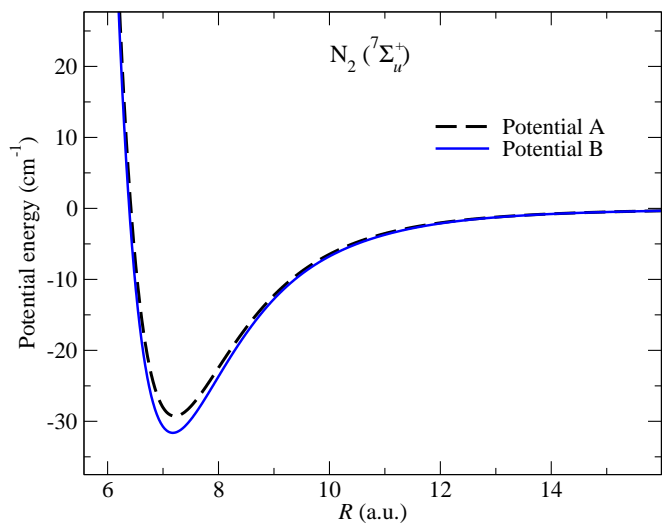}
	\renewcommand{\figurename}{Fig.}
	\caption{ {\it Ab initio} interaction potentials for the $^7\Sigma_u^+$ electronic state of N$_2$ calculated as functions of the internuclear distance. Dashed line -- potential A computed using the UCCSD[T] method, full line -- potential B computed using the MRCCSDT method.}
\label{fig:potentials}
\end{figure}

  For each $R$ point we performed extrapolation to the CBS limit using AVTZ, AVQZ, AV5Z and AV6Z interaction energies. To fit the series of interaction energies we used the empirical formula $E_X=E_{CBS}+ Ae^{-(X-1)}+Be^{-(X-1)^2}$ suggested by Peterson and coworkers \cite{peterson:94,feller:00}, where $X=3,4,5,6$ is the number of ``zetas'' in the basis set. The resulting UCCSD(T)/CBS PEC (labeled as potential A) is shown in Fig.~\ref{fig:potentials}. The potential A has a minimum at $R_e=7.21$ $a_0$ with a well depth of $D_e=29.3$ cm$^{-1}$.  

The $^7\Sigma_u^+$ state has a large multiplicity, leading one to expect a significant contribution of higher excitations in the CCSD(T) method. In order to estimate this contribution, we included the full iterative triple excitations in our {\it ab initio} calculations of the interaction energy for the $^7\Sigma_u^+$ state. The inclusion of full connected triple excitations makes the {\it ab initio} calculations much more computationally demanding, and we employed a single correlation-consistent AVTZ basis set with an additional set of $3s3p2d2f1g$ bond functions (BF) placed at the middle of the N$_2$ bond to reduce computational costs. To perform the full MRCCSDT calculations, we used the MRCC program  \cite{mrcc} by K\' {a}llay {\em et al.} \cite{kallay} interfaced with the MOLPRO code \cite{molpro}. The MRCCSDT calculation used a single-determinant RHF wave function as a reference. The resulting MRCCSDT/AVTZ+BF PEC (labeled as "Potential B") is shown in Fig.~\ref{fig:potentials}. The minimum of the potential B has a well depth of $D_e=31.6$ cm$^{-1}$ and is located at $R_e=7.18$ $a_0$. These values may be compared with the previous {\it ab initio} results $R_e=7.5$ $a_0$ and $D_e=21$ cm$^{-1}$ obtained by Partridge {\it et al.} \cite{Partridge7Sigma}. We note that with the same AVTZ+BF basis the calculated well depth at the UCCSD(T) level is similar to that of potential A. The inclusion of the full triple excitations thus increases the well depth by approximately 7\%. 

To compare the asymptotic behavior of potentials A and B, we fitted their long-range parts between $R=20a_0$ and $R=40a_0$ to the form $-C_6/R^6 - C_8/R^8$. The fitting procedure yields $C_6=23.36$ $E_h a_0^6$ for potential A and $C_6 = 24.0$ $E_h a_0^6$ for potential B. These results are in close agreement with the highly accurate value of $24.2$ \cite{ChuDalgarno}, thereby attesting to the accuracy of our {\it ab initio} calculations. 

Table II lists the bound levels of $^{14}$N$_2(^7\Sigma_u^+)$ and $^{15}$N$_2(^7\Sigma_u^+)$  calculated using potentials A and B in the absence of a magnetic field. Both potentials are deep enough to support three bound levels with $v=0$, 1, and 2. The number of rotational levels decreases from 10 (or 11 for potential B) for $v=0$ to two for $v=2$. We note the presence of accidental degeneracies between the rotational levels corresponding to different $v$: The levels $v=0,\ell=11$ and $v=1, \ell=6$ calculated with potential A have very similar binding energies of $-0.221$ cm$^{-1}$.

\subsection{Scattering calculations}

The Hamiltonian for two S-state atoms such as N($^4{S}_{3/2}$) colliding in the presence of a uniform magnetic field of strength $B$ may be written ($\hbar=1$) 
\begin{equation}\label{H}
\hat{H} = -\frac{1}{2\mu R^2} \frac{\partial^2}{\partial R^2}R + \frac{\hat{\ell}^2}{2\mu R^2} + \hat{V}^\text{sd}(R) + \hat{V}^\text{dip}(R) + \hat{H}_\text{A} + \hat{H}_\text{B} ,
\end{equation}
where $\nu=\text{A}\,, \text{B}$ enumerates the atoms, $\mu$ is the reduced mass of the N$_2$ molecule, $\hat{\ell}$ is the orbital angular momentum for the collision, and  $R=|{\bm R}|$ is the interatomic separation. The Hamiltonian of the isolated atom $\nu$ is given by
\begin{equation}\label{Hat}
\hat{H}_\nu = \gamma_\nu \hat{I}_\nu\cdot\hat{S}_\nu + 2\mu_0 B\hat{S}_{\nu_z} - \frac{\mu_{{I}_\nu}}{I_\nu} \hat{I}_{\nu_z},
\end{equation}
where $\hat{S}_\nu$ and $\hat{I}_\nu$ are the electron and nuclear spins, $\mu_0$  is the Bohr magneton, $\gamma_\nu$ is the hyperfine constant, and $\mu_{I\nu}$ is the nuclear magnetic moment. In this work, we consider both naturally occurring nitrogen isotopes, fermionic $^{14}$N ($I_\nu=1/2$, $\gamma/2\pi=10.451$ MHz) and bosonic $^{15}$N ($I_\nu=1$, $\gamma/2\pi=14.646$ MHz) \cite{Pipkin}. The operators $\hat{S}_{\nu_z}$ and $\hat{I}_{\nu_z}$ yield the projections of $\hat{S}_\nu$ and $\hat{I}_\nu$ on the space-fixed quantization axis defined by the external magnetic field. The magnetic dipole-dipole interaction is \cite{Verhaar}
\begin{equation}\label{Vdip}
\hat{V}^\text{dip} = -\left( \frac{24\pi}{5} \right)^{1/2} \frac{\alpha^2}{R^3} \sum_{q=-2}^{2} (-)^q Y_{2-q}(\hat{R}) [\hat{S}_\text{A}\otimes \hat{S}_\text{B}]^{(2)}_q,
\end{equation}
where $\alpha$ is the fine-structure constant, $[\hat{S}_\text{A}\otimes \hat{S}_\text{B}]$ is a second-rank tensor product of atomic spin operators, and $Y_{kq}(\hat{R})$ are the spherical harmonics. The vector $\hat{R}={\bm R}/R$ describes the orientation of the N$_2$ collision pair in the space-fixed coordinate frame.

 The spin-dependent interaction potential between the atoms may be written as \cite{Verhaar}
\begin{equation}\label{Vsd}
\hat{V}^\text{sd}(R) = \sum_{S=|S_\text{A}-S_\text{B}|}^{S_\text{A} + S_\text{B}} \sum_{M_S=-S}^S V_S(R) |SM_S \rangle \langle SM_S|,
\end{equation}
where  $\hat{S}=\hat{S}_\text{A} + \hat{S}_\text{B}$ is the total spin of the collision complex and $M_S=\MSA+\MSB$ is the projection of $\hat{S}$ on the space-fixed quantization axis. In this work, we use the accurate {\it ab initio} interaction potentials for the $^7\Sigma_u^+$ electronic state of N$_2$ ($S=3$) calculated as described in Sec. IIIA.

Equation (\ref{Vsd}) is parametrized by four spin-dependent interaction potentials of the N$_2$ molecule correlating with the lowest dissociation limit N($^4S_{3/2}$) + N($^4S_{3/2}$). In addition to the high-spin $^7\Sigma_u^+$ potential described above, the low-spin electronic states of $^5\Sigma_g^+$, $^3\Sigma_u^+$, and $^1\Sigma_g^+$ symmetries ($S=2$, 1, and 0) should be taken into account. While the $X^1\Sigma_g^+$ and $A'^5\Sigma_g^+$ electronic states were subject to several theoretical studies \cite{N2X,Partridge5Sigma}, no high-quality {\it ab initio} calculations are available for the $A^3\Sigma_u^+$ electronic state. In the absence of more accurate information, we choose to parametrize the spin-dependent interaction (\ref{Vsd}) by the Heisenberg Hamiltonian \cite{ZS,Cote,SC}  
\begin{equation}\label{Vsd_param}
\hat{V}^\text{sd}(R) = V^\text{si}(R) - 2J(R) \hat{S}_\text{A}\cdot\hat{S}_\text{B},
\end{equation}
where $V^\text{si}$ is a spin-independent interaction potential, and $J(R)$ is the spin-exchange (SE) coupling strength. It follows from Eq. (\ref{Vsd_param}) that the interaction potentials for the spin states $S$ and $S-1$ differ exactly by twice the SE coupling strength
\begin{equation}\label{Vdiff}
V_S(R) - V_{S-1}(R) = 2SJ(R).
\end{equation}

Equation (\ref{Vdiff}) allows us to obtain the four PECs in Eq. (\ref{Vsd}) in terms of two parameters: (i) the potential energy curve for the $^7\Sigma_u^+$ state calculated in Sec. IIIA and (ii) the SE coupling strength $J(R)$. For the latter, we use the expression derived by Smirnov and Chibisov \cite{SC,ZS,Cote} 
\begin{equation}\label{SC}
J(R) = C R^\alpha e^{-\beta R},
\end{equation}
where $\beta=\sqrt{8I}$ and $\alpha=7/\beta-1$ are expressed via the atomic ionization energy $I$ ($0.53412$ $E_h$ for the N atom): $\beta=2.0671$ $a_0^{-1}$ and $\alpha=2.3864$. In Sec. IIIC, we will use $C$ as a free parameter to vary the magnitude of the SE coupling in order to explore the sensitivity of scattering cross sections to the interaction potential.



If the weak hyperfine interaction of $^{14}$N is neglected (see Appendix B for a justification), the wave function of the N$_2$ collision complex can be expanded in direct products of electronic spin functions and partial wave states
\begin{equation}\label{Psi}
\Psi = R^{-1}\sum_{\MSA,\MSB} \sum_{\ell, m_\ell} F_{\MSA \MSB \ell m_\ell} (R) \phi^\eta_{\MSA\MSB \ell m_\ell} (\hat{R})
\end{equation}
where
\begin{equation}\label{SymmBasis}
\phi^{\eta}_{\MSA\MSB\ell m_\ell} = \frac{1}{[2(1+\delta_{\MSA \MSB})]^{1/2}} [|\SA\MSA\rangle |\SB\MSB\rangle +\eta(-)^\ell |\SB\MSB\rangle |\SA\MSA\rangle]  |\ell m_\ell\rangle,
\end{equation}
$|S_\nu M_{S_\nu}\rangle$ are the electronic spin basis functions of individual atoms A and B, $M_{S_\nu}$ are the projections of $S_\nu$ on the space-fixed quantization axis, and $|\ell m_\ell\rangle = Y_{\ell m_\ell}(\hat{R})$. The direct-product basis (\ref{SymmBasis}) is properly ordered ($\tA\ge \tB$) and symmetrized to account for the quantum statistics of indistinguishable bosons ($^{15}$N, $\eta= 1$, odd $\ell$) or fermions ($^{14}$N, $\eta=-1$, even $\ell$). 

The matrix elements of the spin-dependent interaction potential (\ref{Vsd}) in the symmetrized basis are 
\begin{multline}\label{symm_matrix_element}
\langle \phi^\eta_{\MSA \MSB \ell m_\ell} | \hat{V}^\text{sd}|\phi^\eta_{\MSA' \MSB'\ell' m_\ell'} \rangle =
 \frac{\delta_{\ell\ell'}\delta_{m_\ell m_\ell'} }{[(1+\delta_{\MSA \MSB})(1+\delta_{\MSA' \MSB'})]^{1/2}}   \\ \times [\langle \SA \MSA | \langle \SB \MSB| \hat{V}^\text{sd}  | \SA\MSA'\rangle |\SB \MSB'\rangle  +\eta(-)^\ell \langle \SA \MSA | \langle\SB \MSB| \hat{V}^\text{sd} | \SB \MSB' \rangle |\SA \MSA' \rangle ].
\end{multline}
The second term on the right-hand side (which arises from symmetrization) can be obtained from the first term by interchanging the indices $\MSA'\leftrightarrow\MSB'$. We therefore only need to evaluate the unsymmetrized matrix element
\begin{equation}\label{matrix_element}
V^\text{sd}_{\MSA \MSB \ell m_\ell; \MSA' \MSB' \ell'  m_\ell'} = \delta_{\ell\ell'}\delta_{m_\ell m_\ell'}  \langle \SA \MSA| \langle \SB \MSB | \hat{V}^\text{sd} | \SA \MSA' \rangle | \SB \MSB'\rangle.
\end{equation}
where the subscripts $\SA$, $\SB$ and $\SA'$, $\SB'$ have been omitted for clarity. Expanding the product of two spin functions in a Clebsh-Gordan series \cite{Zare,Li}, we obtain
\begin{multline}\label{me_Vsd}
V^\text{sd}_{\MSA \MSB \ell m_\ell; \MSA' \MSB' \ell'  m_\ell'} = \delta_{\ell\ell'}\delta_{m_\ell m_\ell'}  \sum_{S,\,M_S} (2S+1) (-)^{M_S}  \threejm{\SA}{\MSA}{\SB}{\MSB}{S}{-M_S}   \\ \times \threejm{\SA}{\MSA'}{\SB}{\MSB'}{S}{-M_S} V_S(R),
 \end{multline}
Because the spin-dependent interaction potential (\ref{Vsd}) is diagonal in the total spin $\hat{S}$ and its space-fixed projection $M_S$, the matrix elements between the fully spin-polarized initial state $|\SA, \MSA = \SA\rangle |\SB,\MSB=\SB\rangle$ and all other spin states vanish identically. Thus, in the absence of the magnetic dipole interaction (see below), spin-polarized atoms can only undergo elastic scattering.

The matrix elements of the magnetic dipole interaction can be derived as described elsewhere \cite{Verhaar,BernardAlex,Hutson}. Here, we only present the final result
 \begin{multline}\label{me_Vsd}
V^\text{dip}_{\MSA \MSB \ell m_\ell; \MSA' \MSB' \ell'  m_\ell'}  =  -\frac{\sqrt{30}\alpha^2}{R^3} \sum_q (-)^{m_\ell+\SA+\SB - M_S} [(2\ell+1)(2\ell'+1)]^{1/2} \\ \times [(2\SA+1)\SA(\SA+1)]^{1/2} [(2\SB+1)\SB(\SB+1)]^{1/2} \threejm{\ell}{0}{2}{0}{\ell'}{0}\threejm{\ell}{-m_\ell}{2}{-q}{\ell'}{m_\ell'} \\ \times \sum_{q_\text{A},q_\text{B}}  \threejm{1}{q_\text{A}}{1}{q_\text{B}}{2}{-q} \threejm{\SA}{-\MSA}{1}{q_\text{A}}{\SA}{\MSA'} \threejm{\SB}{-\MSB}{1}{q_\text{B}}{\SB}{\MSB'},
\end{multline}
Unlike the interaction potential (\ref{Vsd}), the magnetic dipole interaction does not conserve $M_S$ and couples the fully spin-stretched state $|\SA, \MSA = \SA\rangle |\SB,\MSB=\SB\rangle$ to other spin states, thereby inducing spin-flipping transitions. By transforming Eq. (\ref{me_Vsd}) to the total spin representation, one can show that the matrix elements of the magnetic dipole interaction vanish identically unless $S-S'=0,\pm 2$ \cite{ZoranCr, BernardAlex}. Thus, the mechanism of dipolar relaxation in collisions of fully spin-polarized N atoms is completely determined by the electronic states of $^7\Sigma_u^+$ and $^3\Sigma_u^+$ symmetries \cite{ZoranCr}. This mechanism is typical of light $S$-state atoms \cite{note}. The rate constants for dipolar relaxation can be measured by observing collision-induced loss of atoms from a magnetic trap as described in Sec. II.

A system of close-coupled (CC) equations for the radial functions $F_{\MSA\MSB \ell m_\ell}(R)$ results when the expansion (\ref{Psi}) is combined with the Schr{\"o}dinger equation with Hamiltonian (\ref{H}). The CC equations are integrated numerically on a radial grid extending from $R=4.0$ $a_0$ to $R=$ 50.0 $a_0$ with a constant spacing of 0.04 $a_0$ using the improved log-derivative algorithm \cite{LogDerivative}. The calculations are carried out separately for each total angular momentum projection $M=\MSA+\MSB+m_\ell$, which is conserved for collisions in external fields. 
 The scattering $S$-matrix is computed directly in the uncoupled basis (\ref{SymmBasis}) and used to evaluate the cross sections for collision-induced transitions between different Zeeman states \cite{Burke,Roman2004,NJP}
\begin{multline}\label{CrossSection}
\sigma_{\MSA \MSB \to \MSA'\MSB'}= \frac{\pi (1+\delta_{\MSA\MSB})}{k_{\MSA\MSB}^2} \sum_{M} \sum_{\ell, m_\ell} \sum_{\ell',m_\ell'} | \delta_{\MSA\MSA'}\delta_{\MSB\MSB'}\delta_{\ell\ell'}\delta_{m_\ell m_\ell'} \\ -S_{\MSA\MSB\ell m_\ell; \MSA'\MSB'\ell'm_\ell'}^{M}  |^2,
\end{multline}
where the factor $(1+\delta_{\MSA\MSB})$ accounts for the indistinguishability of colliding atoms. In order to make sure that our numerical results are correct, we repeated scattering calculations with a different code \cite{ZoranCr} and obtained the same results.  A total of 64 channels were coupled for $M=0$, corresponding to the partial wave states with $\ell < \ell_\text{max} = 13$. The calculated elastic and inelastic cross sections were converged to better than 3\% with respect to the initial and final integration distances and $\ell_\text{max}$ for all collision energies and magnetic fields studied in this work.

By averaging the symmetrized cross sections (\ref{CrossSection}) over a Maxwell-Boltzmann distribution of collision energies in the interval $E_C=0.01-2$ cm$^{-1}$ with a grid step size of 0.01 cm$^{-1}$, we obtain state-resolved dipolar relaxation rates as functions of temperature $T$
\begin{equation}\label{MB}
K_{\MSA \MSB \to \MSA'\MSB'}(T)= \left(\frac{8\beta}{\pi\mu}\right)^{1/2} \int \beta E_C \sigma_{\MSA\MSB\to \MSA'\MSB'} (E_C) e^{-E_C/k_BT}dE_C,
\end{equation}
where $\beta=1/k_BT$ and $k_B$ is Boltzmann's constant.

\subsection{Theoretical results}

\subsubsection{Cross sections and rate constants for dipolar relaxation}

Figure \ref{fig:Nlevels} shows the energy levels of $^{14}$N as functions of the applied magnetic field. At zero field, the ground state is split by the hyperfine interaction into three levels with $F=5/2$, 3/2, and 1/2. The inset of Fig.~\ref{fig:Nlevels} shows the hyperfine splittings as functions of the applied magnetic field. As shown in Appendix A, the hyperfine splittings have a minor effect on N~+~N collisions except at very small magnetic fields ($<$20 G), so it is a good approximation to consider bare spin states $|\SA,\MSA\rangle$ with $\MSA = -3/2, \ldots\,3/2$. The magnetic trapping experiments described in Sec. II select N atoms in the fully spin-polarized state $|S=M_S=3/2\rangle$, so in the following we will only consider collisions of N atoms initially in this state.

The calculated cross sections for elastic scattering and dipolar relaxation in $^{14}$N + $^{14}$N and $^{15}$N + $^{15}$N collisions are displayed in Fig.~\ref{fig:NsigmavsE} as functions of collision energy  $E_C$. In the limit of vanishing collision energy, the variation of the cross sections with $E_C$ is determined by the Wigner threshold law; the cross sections for elastic scattering vary as $E_C^{2\ell}$ and those for dipolar relaxation as $E_C^{\ell-1/2}$. Collisions of identical fermions such as $^{14}$N are determined by $p$-wave scattering, so both the elastic and inelastic cross sections vanish as $E_C$ tends to zero. The situation for the bosonic isotope $^{15}$N is the opposite: the inelastic cross section diverges and the elastic cross section approaches a constant. At collision energies above $\sim$0.01 cm$^{-1}$, partial waves with $\ell>1$ begin to factor in, altering the dependence of the cross sections of collision energy, and leading to the appearance of shape resonances. A particularly pronounced resonance at $E_C\sim 1.3$ cm$^{-1}$ corresponds to the quasibound $^{14}$N$_2$($^7\Sigma_u^+$) molecule in the $\ell=7$ rotational state.

In Fig. \ref{fig:NsigmavsEvsB}, we plot the cross sections for dipolar relaxation as functions of collision energy and magnetic field. Magnetic fields increase the splitting between the incoming and outgoing collision channels and suppress dipolar relaxation \cite{BernardAlex}. The suppression is quite pronounced at high magnetic fields (on order of 1 T), but does not alter the dependence of the cross sections on collision energy. In particular, increasing the magnetic field from 0.01 T to 1 T reduces the lifetime of the $\ell=7$ shape resonance by more than two orders of magnitude, but leaves its position intact. The same is true for the low-energy $\ell=2$ shape resonance in $^{15}$N. As shown in Fig. \ref{fig:NsigmavsEvsB} for both $^{14}$N and $^{15}$N, the decrease of the cross sections with $B$ is not always monotonous.

As discussed in the Introduction, collisional stability is a key ingredient to efficient use of trapped atomic gases for sympathetic cooling of molecular ensembles. Fig. \ref{fig:NratevsT}(a) shows the rate constants for dipolar relaxation calculated by thermally averaging the cross sections shown in Fig.~\ref{fig:NsigmavsE} at a fixed magnetic field of 0.1 T. The relaxation rates for both N isotopes display broad maxima at $T\sim 10$ mK (for $^{14}$N) and $T\sim 50$ mK (for $^{15}$N). The rate for $^{14}$N features an additional maximum near 1 K due to the $\ell=7$ shape resonance shown in Fig.~\ref{fig:NsigmavsE}. At temperatures below 5 mK, inelastic collisions occur in the Wigner $s$-wave regime, and the rate constants for dipolar relaxation tend to zero for $^{14}$N and approach a constant value of $5.5\times 10^{-13}$ cm$^3$/s for $^{15}$N. The ratios of the rate constants for elastic scattering and dipolar relaxation displayed in Fig. \ref{fig:NratevsT}(b) remain high ($\gamma >100$) down to $\sim$10 mK for $^{14}$N and $\sim$2 mK for $^{15}$N. This result shows that trapped ensembles of $^{14}$N and $^{15}$N atoms with densities 10$^{12}$ cm$^{-3}$ will have lifetimes $\sim$2 s over a wide range of temperatures from 1 mK to 1 K. We note that because the elastic cross section for $^{14}$N  becomes very small at $T<1$ mK, $^{15}$N is a more promising candidate for cooling molecules to temperatures below 1 mK, whereas both $^{14}$N and $^{15}$N isotopes appear suitable for sympathetic cooling to temperatures above 1 mK.


Figure \ref{fig:NratevsB} shows the temperature dependence of state-to-state rate constants for dipolar relaxation in $^{14}$N + $^{14}$N collisions (\ref{MB}). The single spin-flip transition $|\MSA=3/2\rangle |\MSB=3/2\rangle \to |\MSA'=1/2\rangle |\MSB'=3/2\rangle$ dominates over the whole range of magnetic fields at both 0.1 K and 0.6 K, and the double spin-flip transition $|\MSA=3/2\rangle |\MSB=3/2\rangle \to |\MSA'=1/2\rangle |\MSB'=1/2\rangle$ is the next most efficient. The rate constants for other transitions (changing $M_S$ by 2 or more) are negligibly small.





\subsubsection{Sensitivity of trap loss rates to the interaction potential}

In order to verify the reliability of our scattering calculations, it is essential to analyze various sources of uncertainty that can affect the accuracy of numerical results for the dipolar relaxation cross sections and trap loss rates. In addition to numerical convergence (Sec. IIIB) and the neglect of the hyperfine interaction (Appendix B), we consider two additional sources of uncertainty, of which the first arises from neglecting the SE interaction [setting $C=0$ in Eq. (\ref{SC})] and the second from inaccuracies in the {\it ab initio} interaction potential for the $^7\Sigma_u^+$ state of N$_2$ calculated in Sec. IIIA.

To examine the sensitivity of the calculated dipolar relaxation rates to the SE interaction, we calculated the rates (\ref{Kloss}) as functions of the SE parameter $C$ (\ref{SC}). To estimate the range of variation of $C$, we used the {\it ab initio} results available for the $^5\Sigma_g^+$ electronic state. Figure~\ref{fig:PotentialsvsC} shows the PEC for the $^5\Sigma_g^+$ state of N$_2$ calculated for selected values of $C$ using Eqs. (\ref{Vdiff}) and (\ref{SC}). In the absence of the SE interaction, the $^5\Sigma_u^+$ potential is identical to the $^7\Sigma_u^+$ potential. As $C$ increases, the $^5\Sigma_u^+$ potential becomes deeper and shifts towards smaller $R$. While the Heisenberg exchange Hamiltonian (\ref{Vsd}) cannot accurately describe the shape of the {\it ab initio} PEC for the $^5\Sigma_g^+$ state \cite{Partridge5Sigma}, the long-range part of the curve is fairly well reproduced at $C=0.3$ $E_h$. Based on the comparison presented in Fig.~\ref{fig:PotentialsvsC}, we choose to vary $C$ in the range from 0 to 0.5 $E_h$ with a grid step of 0.05 $E_h$.

As shown in Fig. \ref{fig:NratevsC}, the rate constants for dipolar relaxation (\ref{Kloss}) do not vary strongly with $C$, except at the lowest temperature studied (0.1 K). The dependence of the calculated rates on $C$ is not monotonous, and the largest deviation from $C=0$ values used as a reference in this work does not exceed 15 \%. A similar lack of sensitivity has recently been observed in quantum calculations of dipolar relaxation in collisions of spin-polarized Eu atoms \cite{YS}. We emphasize, however, that the SE interaction in the Eu$_2$ collision complex is several orders of magnitude weaker than in N$_2$, so the range of SE splittings probed in Ref. \cite{YS} was much narrower than explored in this work. The results presented in Fig. \ref{fig:NratevsC} therefore suggest that the rate constants for dipolar relaxation are insensitive to much {\it larger} variations of the SE interaction (on the order of several eV).

To understand how the calculated inelastic rates are affected by the uncertainties in the $^7\Sigma^+_u$ interaction potential, we scaled the potential by a constant factor $\lambda$ and calculated the temperature dependence of $\langle K_\text{in}\rangle$ for 20 equally spaced values of $\lambda$ in the interval 0.9 -- 1.1 ($\lambda=1$ corresponds to unscaled potential B). The relatively small range of $\lambda$ chosen reflects the high level of accuracy of the {\it ab initio} interaction potential presented in Sec. IIIA ($<$10\%).  Figure \ref{fig:Nratevslambda} shows the dependence of $\langle K_\text{in}(\lambda)\rangle$ for three selected temperatures. At $T=0.6$ K, the variation of $\langle K_\text{in}\rangle$ is within 15 \% for the whole range of $\lambda$, demonstrating that our results are robust against both lessening ($\lambda<1$) and deepening ($\lambda>1$) of the interaction potential. The sensitivity of the calculated relaxation rates to $\lambda$ increases at low temperatures, reaching a maximum at $T=0.1$~K. This is an expected result since the variation of the cross section with $\lambda$ should be most pronounced in the ultracold limit, where the $s$-wave scattering cross section exhibits a resonance-like variation as a function of $\lambda$ \cite{Hutson07}.


Figure \ref{fig:NratevsTemp} shows the temperature dependence of the calculated dipolar relaxation rates for $^{14}$N + $^{14}$N collisions. The error bars represent maximum possible deviations from the mean value of $\langle K_\text{in}\rangle$ defined as the value calculated in the absence of the SE interaction ($C=0$) for unscaled potential B ($\lambda=1$). We evaluate the error bars by finding the extrema of the calculated functions $\langle K_\text{in}(\lambda)\rangle$ and $\langle K_\text{in}(C)\rangle$ for each $T$. The rate constants decrease and the error bars shrink with increasing temperature. The results presented in Table III indicate that while imperfections in the $^7\Sigma_u^+$ interaction potential are the dominant source of uncertainty at temperatures below 0.6~K, omission of the SE interaction introduces the same amount of error at $T = 0.6$~K and becomes the major source of uncertainty above this temperature. From Figs. \ref{fig:NratevsC} and \ref{fig:Nratevslambda}, we observe that scaling the interaction potential tends to increase the inelastic rates, whereas varying the strength of the SE interaction does not always lead to the monotonous variation. As a result, uncertainties in the interaction potential determine the upper error bar at $T\le 0.6$~K, and those in the SE interaction determine the lower error bar at all temperatures.

Figure \ref{fig:Eratevslambda} shows the variation of the elastic collision rates for $^{14}$N + $^{14}$N with $\lambda$. Strong sensitivity to $\lambda$ is apparent over the whole temperature range. 
As discussed in Sec. IIC, the calculated ratio of the rate constants for elastic scattering and dipolar relaxation in $^{14}$N~+~$^{14}$N collisions is consistent with the measured value of $\gamma$ shown in Fig.~\ref{fig:gamma}.

We note that the uncertainties arising from lack of knowledge of the SE interaction can be reduced by performing calculations with the accurate {\it ab initio} interaction potential for the $^3\Sigma_u^+$ electronic state of N$_2$, that is, using the exact Eq. (\ref{Vsd}) instead of the approximate Heisenberg parametrization (\ref{Vsd_param}). Such calculations are currently in progress \cite{JacekTBP}, and preliminary results show that the calculated relaxation rates are very close (within 5\%) to the results obtained with the SE interaction omitted, thereby lending support to our claim that the major source of uncertainty in the calculated dipolar relaxation rates comes from the interaction potential for the $^7\Sigma_u^+$ electronic state, rather than the SE interaction.

\section{Summary}

We have presented a combined experimental and theoretical study of collisional properties of cold spin-polarized atomic nitrogen. We have trapped large numbers of $^{14}$N and $^{15}$N atoms for tens of seconds and measured their dipolar relaxation rates at 600 mK. Based on these measurements and theoretical calculations of trap loss rates, we have determined the number density of trapped N atoms to be $(5\pm2) \times 10^{11}$~cm$^{-3}$. 

Our theoretical analysis of dipolar relaxation in N + N collisions is based on accurate {\it ab initio} interaction potentials for the $^7\Sigma_u^+$ state of N$_2$ computed using highly correlated coupled cluster methods (Sec. IIIA). By solving the multichannel scattering problem, we obtained the cross sections and rate constants for dipolar relaxation in N + N collisions over a wide range of collision energies and magnetic fields (Sec. IIIB). The calculated relaxation rates for both $^{14}$N and $^{15}$N isotopes are similar in the multiple partial wave regime ($T>5$ mK), but display a very different behavior at ultralow  temperatures due to the effects of quantum statistics (Figs. \ref{fig:NsigmavsEvsB} and \ref{fig:NratevsT}). The rate constants for dipolar relaxation in N + N collisions are on the order of $10^{-13}$ cm$^3$/s, indicating that spin-polarized N atoms are stable against collisional relaxation in the temperature range between 1 mK and 1 K. The results presented in Fig. \ref{fig:NratevsT} indicate that sympathetic cooling of paramagnetic molecules with N atoms will be efficient provided the probabilities for inelastic relaxation in N-molecule collisions are not very large. At $T<1$ mK, the elastic cross section for $^{14}$N + $^{14}$N decreases dramatically and the elastic-to-inelastic ratio for $^{15}$N + $^{15}$N drops below 100 (Fig. \ref{fig:NratevsT}). Thus, neither $^{14}$N nor $^{15}$N appears suitable for sympathetic cooling of molecules below 1 mK. It might be possible to further reduce the temperature of trapped molecules via evaporative cooling at low magnetic fields \cite{NJP} once N atoms are removed from the trap.

In agreement with a recent theoretical study of dipolar relaxation in Eu + Eu collisions \cite{YS}, we found that the calculated rate constants for trap loss in collisions of spin-polarized N atoms are not sensitive to the magnitude of the SE interaction. We identified inaccuracies in the interaction potential for the $^7\Sigma_u^+$ electronic state of N$_2$ as the major source of uncertainty in our theoretical results. These inaccuracies (on the order of 10\%) lead to large variations of the calculated relaxation rates at temperatures below 0.1 K (Fig. \ref{fig:Nratevslambda}), but have a minor effect at the experimental temperature of $0.6$ K (Fig.~\ref{fig:NratevsTemp} and Table III), enabling accurate calibration of the trapped N atom density (Sec. II). 
The results shown in Fig. \ref{fig:NratevsTemp} demonstrate that rigorous quantum scattering calculations based on {\it ab initio} interaction potentials are capable of providing quantitative accuracy required for the interpretation of cold collision experiments in the temperature range between 0.1 and 0.7 K.

The calculated rate constants for dipolar relaxation at temperatures below 0.1 K (Fig.~\ref{fig:NratevsT}) are subject to large uncertainties arising from imperfections in {\it ab initio} interaction potentials. As shown in Fig.~\ref{fig:NratevsTemp}, the calculated trap loss rates at $T=0.1$ K are only accurate to within a factor of 3. At lower temperatures, scattering resonances similar to those shown in Fig.~\ref{fig:NsigmavsE} may have a profound effect on collision dynamics. The positions and widths of these resonances are extremely sensitive to tiny variations in the interaction potentials and hence cannot be predicted quantitatively. As in the case of ultracold collisions of alkali-metal atoms \cite{Julienne},  empirical adjustment of the {\it ab initio} interaction potentials may be required to obtain quantitative agreement with future experimental studies of N + N collisions at temperatures below 0.1 K.

Our findings indicate that spin-polarized nitrogen atoms may have favorable collisional properties over a wide range of temperatures and magnetic fields, making them promising candidates for future experiments on sympathetic cooling of open-shell molecules such as NH \cite{NH-N} to temperatures $\sim$1 mK. A detailed study of cold N + NH collisions in a magnetic trap will be presented in future work \cite{FutureWorkNH-N}. The moderate magnetic moment of N atoms (3$\mu_B$) is large enough to enable efficient magnetic trapping and evaporative cooling \cite{NH-N,MattThesis} and small enough to make collision-induced dipolar relaxation inefficient. The latter property is particularly important since large inelastic loss rates recently observed in collisions of highly magnetic atoms \cite{Cr,Tm} make these atoms unsuitable for sympathetic cooling of molecules in permanent magnetic traps.

\acknowledgements

 This work was supported by NSF (grant No. PHY-0757157), Air Force Office of Scientific Research (grant No. FA9550-07-1-0492), the Chemical Sciences, Geosciences, and Biosciences Division of the Office of Basic Energy Science, Office of Science, U.S. Department of Energy, and NSF grants to the Harvard-MIT Center for Ultracold Atoms and the Institute for Theoretical Atomic, Molecular and Optical Physics at Harvard University and the Smithsonian Astrophysical Observatory.  J. K. acknowledges financial support from NSF (grant No. CHE-0848110) to M. H. Alexander. B. Z. was supported by NSF grant No. PHY-0758140.

\appendix
\section{Rate equations}



Here, we present the derivation of the rate equation (\ref{eq:combined_loss}). Assuming that the process of trap loss is irreversible, the time decay of trapped N atom density $n$ due to dipolar relaxation in binary N + N collisions can be described by the following rate equation \cite{Julienne,Burke}
\begin{equation}\label{RateEquation}
-\dot{n} = \sum_{\MSA',\MSB'} w_{\MSA'\MSB'} K_{{\textstyle\frac{3}{2}\frac{3}{2}} \to \MSA'\MSB'}(B,T) n^2,
\end{equation}
where $K_{ {\textstyle\frac{3}{2}\frac{3}{2}}  \to \MSA'\MSB'}(B,T)$ are state-resolved rate constants for dipolar relaxation (\ref{MB}). The weighting factors in Eq. (\ref{RateEquation}) serve to distinguish between single spin-flip collisions, in which only one atom is lost ($w_{ \frac{1}{2}\frac{3}{2} }=1$) and double spin-flip collisions, in which both atoms are lost ($w_{ \frac{1}{2}\frac{1}{2} }=2$). Taking into account only the dominant relaxation channels shown in Fig. \ref{fig:NratevsB}, we can rewrite the rate equation (\ref{RateEquation}) in the form 
\begin{equation}\label{RateEquation2}
-\dot{n} =  K_\text{in}(B,T) n^2,
\end{equation}
where
\begin{equation}\label{Kloss}
K_\text{in}(B,T) = \textstyle{\frac{1}{2}} \bigl{[} 2 K_{\textstyle \frac{3}{2}\frac{3}{2} \to \frac{1}{2}\frac{1}{2}} (B,T) +   K_{\textstyle \frac{3}{2}\frac{3}{2} \to \frac{1}{2}\frac{3}{2}}(B,T) \bigr{]}
\end{equation}
is the total rate constant for trap loss and the factor of 1/2 is introduced to account for indistinguishability of collision partners \cite{Burke}. The right-hand side of Eq. (\ref{Kloss}) can be evaluated in terms of the partial rate constants given by Eq. (\ref{MB}).

The rate constants defined by Eq. (\ref{Kloss}) characterize the dynamics of dipolar relaxation in the presence of a uniform magnetic field. The trapping field generated in our apparatus (Fig. 1) is, however, highly inhomogeneous, so the calculated loss rates (\ref{Kloss}) should be averaged over the magnetic field distribution of the trap. To do this, we assume a trapped sample density distribution of the form 
\begin{equation}\label{TrapDistribution}
n(\mathbf r) = n_0 U({\mathbf r},T)
\end{equation}
where $U({\mathbf r},T) = \exp [-\mu B(\mathbf r ) / k_\mathrm{B} T]$ is the magnetic field distribution of the trap, $n_0$ is the density of N atoms at the trap center, $B(\mathbf r)$ is the trapping field, and $T$ is the atom temperature.  The trapping field is calculated numerically from the known electromagnetic coil profiles, and then fit to an 11 term polynomial \cite{MattThesis}.

Integration of Eq.~(\ref{RateEquation2}) over the trap volume using the density distribution of Eq. (\ref{TrapDistribution}) yields an expression for total trap loss 
\begin{equation}
-\dot{n}_0 =  \frac{\int K_\text{in}(B({\mathbf r}),T) U({\mathbf r},T)^2 dV }{\int U({\mathbf r},T) dV}  n_0^2 = \frac{1}{7.6}\langle K_\text{in} (T)\rangle  n_0^2,
\end{equation}
where 
\begin{equation}\label{Vaveraging}
\langle K_\text{in}(T) \rangle \equiv \frac{\int K_\text{in}(B({\mathbf r}),T) U({\mathbf r},T)^2dV} {\int U({\mathbf r,T})^2 dV} 
\end{equation}
is the average rate constant for trap loss, and the value of $1/7.6$ comes from the numeric evaluation of the expression $\frac{\int U({\mathbf r},T)^2 dV} {\int U({\mathbf r},T) dV}  \approx \frac{1}{7.6}$ for the experimental trap geometry (Fig.~1)~\cite{MattThesis}. 


\section{Hyperfine interaction}

\begin{figure}[t]
	\centering
     \includegraphics[width=0.7\textwidth, trim = 0 0 0 -40]{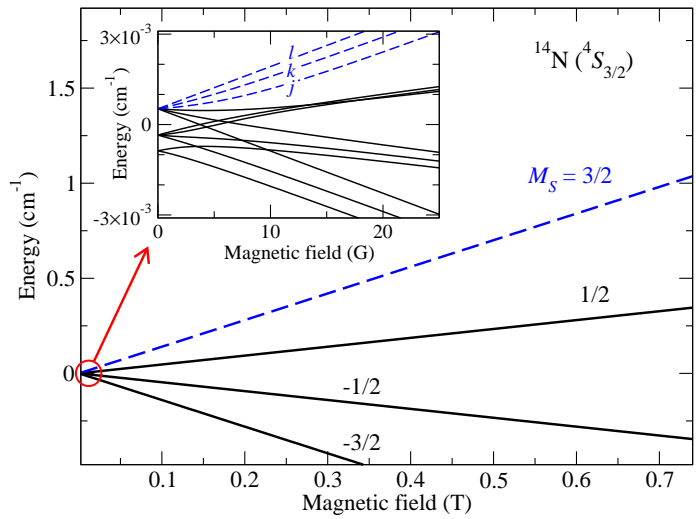}
	\renewcommand{\figurename}{Fig.}
	\caption{Zeeman energy levels of $^{14}$N as functions of the applied magnetic field. The inset shows details of the hyperfine structure at low fields. The states are labeled according to their spin projections $\MSA$. The highest-energy low-field-seeking state $|\SA=3/2,\MSA=3/2\rangle$ is shown by the dashed line.}
	\label{fig:Nlevels}
\end{figure}

\begin{figure}[t]
	\centering
     \includegraphics[width=0.65\textwidth, trim = 0 0 0 -50]{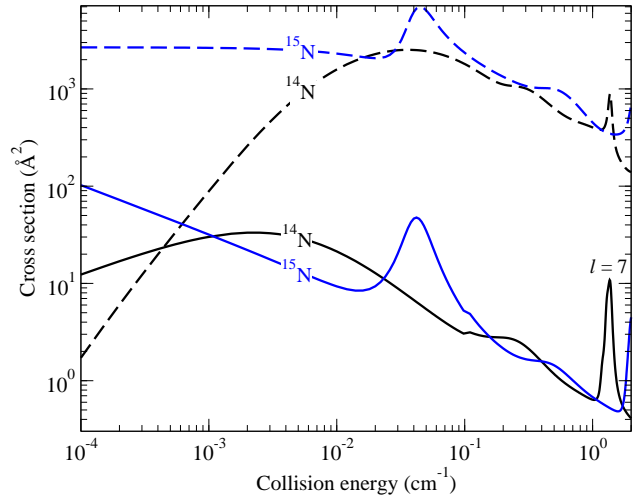}
	\renewcommand{\figurename}{Fig.}
	\caption{Cross sections for dipolar relaxation (full lines) and elastic scattering (dashed lines) in $^{14}$N + $^{14}$N and $^{15}$N + $^{15}$N collisions plotted vs collision energy $E_C$ for $B=0.1$ T and $C=0$. The peak around $E_C=1.3$ cm$^{-1}$ corresponds to an $\ell=7$ shape resonance.}
	\label{fig:NsigmavsE}
\end{figure}

\begin{figure}[t]
	\centering
     \includegraphics[width=1.0\textwidth, trim = 0 100 0 0]{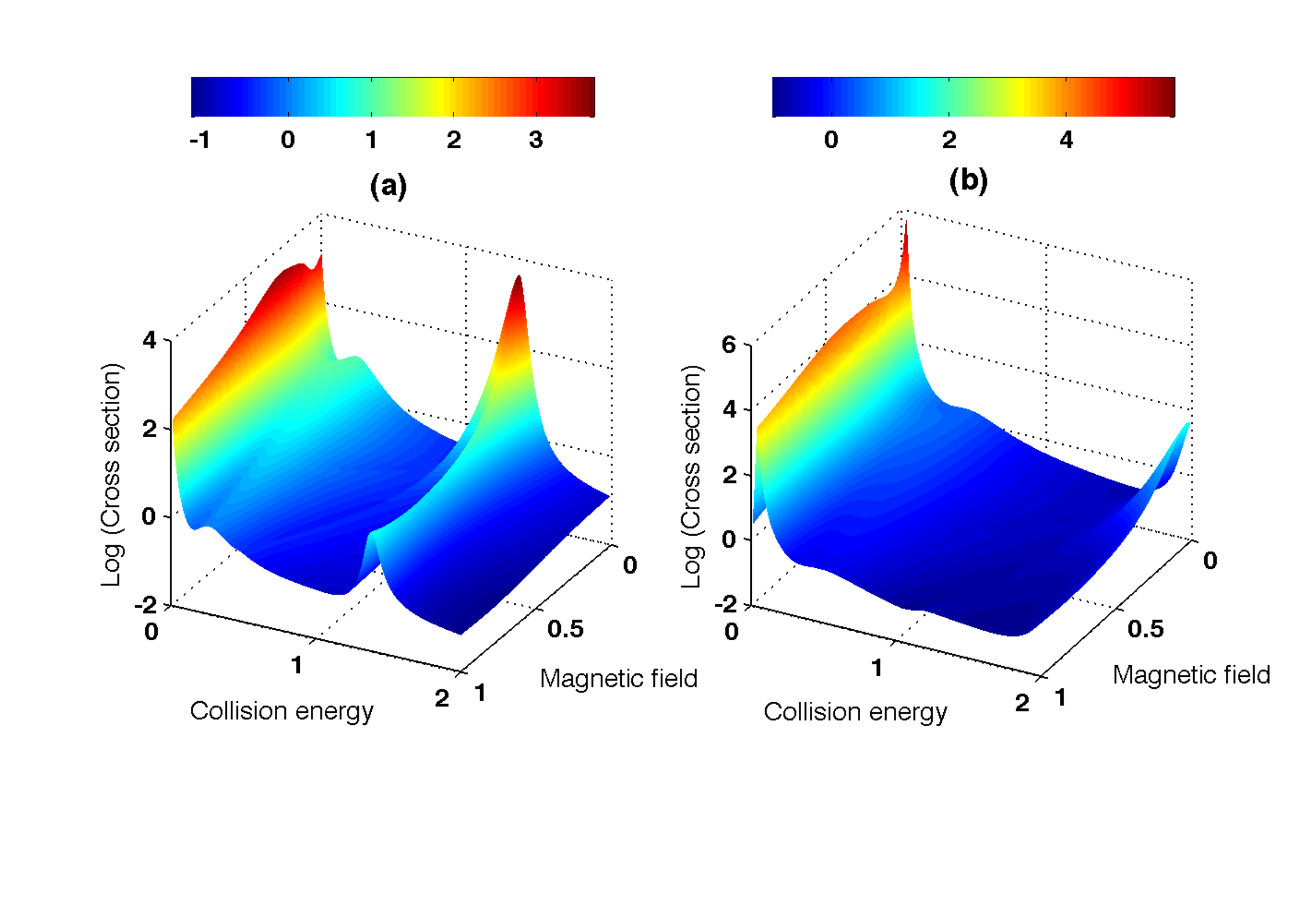}
	\renewcommand{\figurename}{Fig.}
	\caption{(a) Cross sections for dipolar relaxation in $^{14}$N + $^{14}$N collisions as functions of collision energy (in cm$^{-1}$) and magnetic field (in T). Note the presence of the $\ell=7$ shape resonance marked in Fig.~8 and its evolution with magnetic field. (b) Same but for $^{15}$N + $^{15}$N collisions. The cross sections are evaluated for $C=0$. }
	\label{fig:NsigmavsEvsB}
\end{figure}

\begin{figure}[t]
	\centering
     \includegraphics[width=0.65\textwidth, trim = 0 0 0 -50]{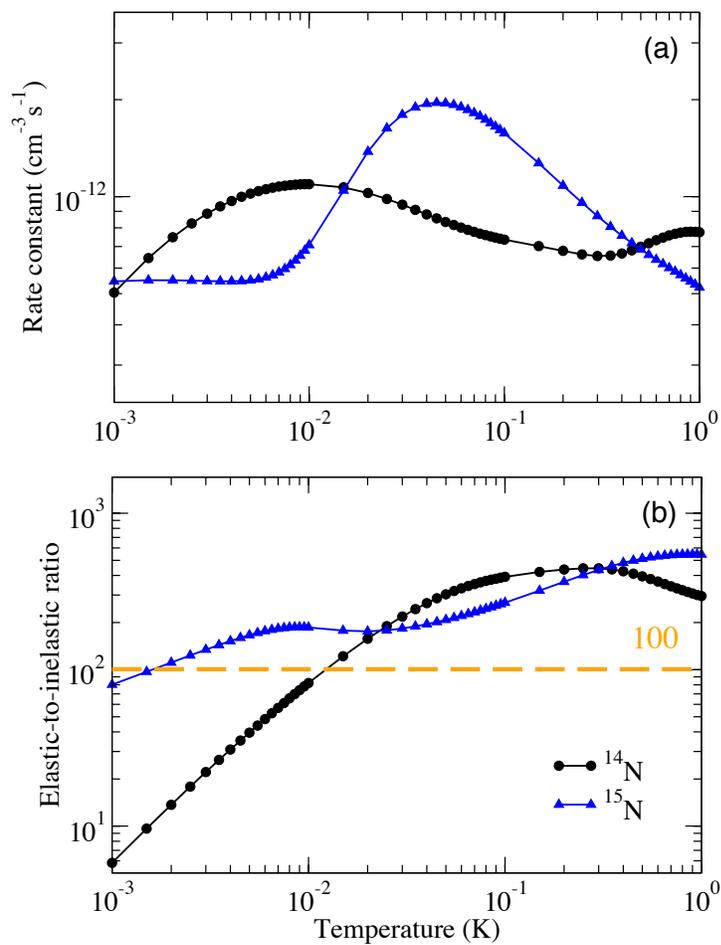}
	\renewcommand{\figurename}{Fig.}
	\caption{(a) Rate constants for dipolar relaxation in collisions of $^{14}$N atoms (circles) and $^{15}$N atoms (triangles) as functions of temperature. (b) Thermally averaged ratios of the rate constants for elastic scattering and dipolar relaxation. The magnetic field is 0.1 T.}
	\label{fig:NratevsT}
\end{figure}

\begin{figure}[t]
	\centering
     \includegraphics[width=0.7\textwidth, trim = 0 0 0 -50]{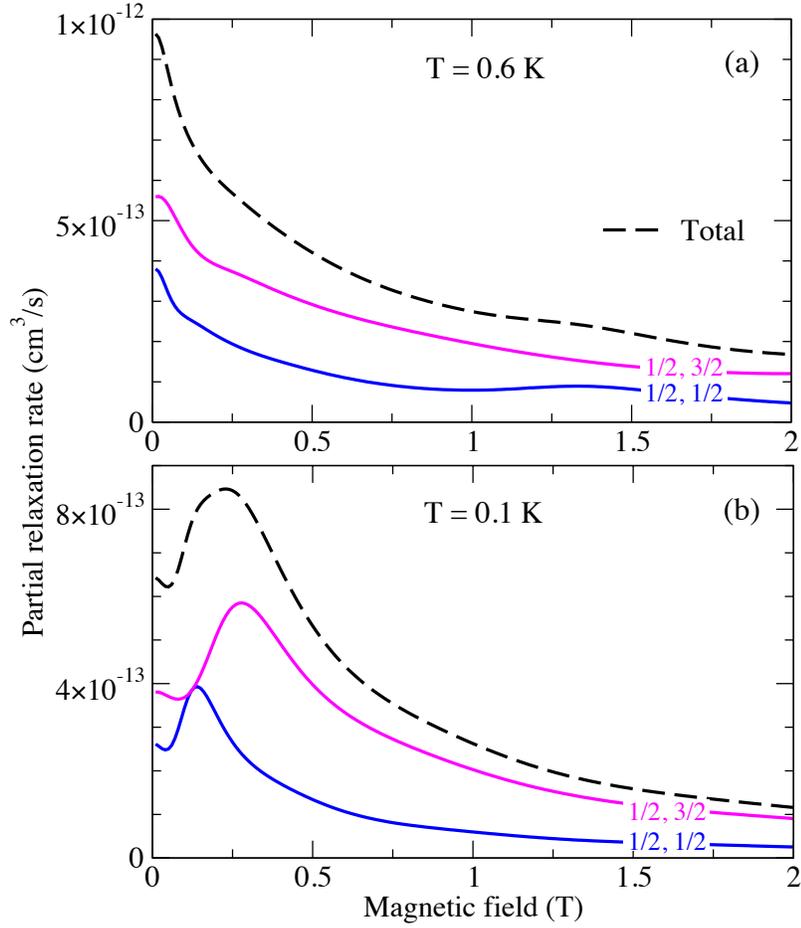}
	\renewcommand{\figurename}{Fig.}
	\caption{(a) Partial rate constants for dipolar relaxation in $^{14}$N + $^{14}$N collisions as functions of magnetic field calculated with potential A for (a) $T=0.6$ K, (b) $T=0.1$ K and $C=0$. The total rate constant for dipolar relaxation is shown by the dashed line.}
	\label{fig:NratevsB}
\end{figure}

\begin{figure}[t]
	\centering
     \includegraphics[width=0.75\textwidth, trim = 0 0 0 0]{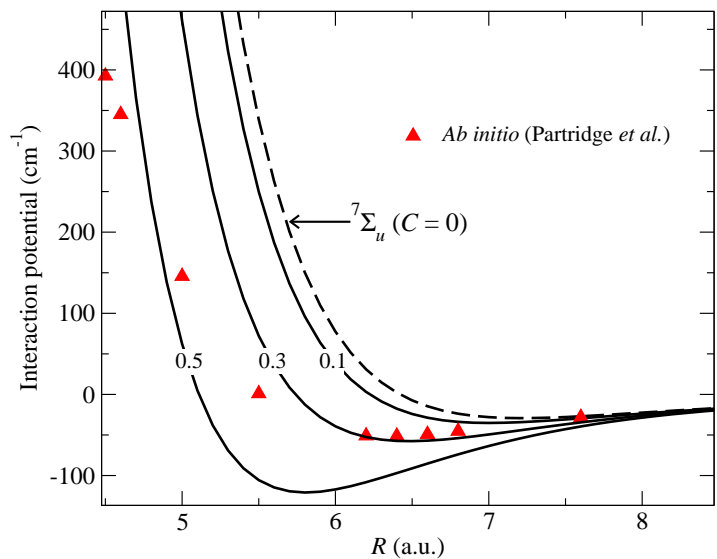}
	\renewcommand{\figurename}{Fig.}
	\caption{Potential energy curve for the $A'^5\Sigma_g^+$ electronic state of N$_2$ calculated for different values of the SE parameter $C$ indicated in the graph. The MRCI+Q results of Partridge {\it et al.} \cite{Partridge5Sigma} are shown by triangles.}
	\label{fig:PotentialsvsC}
\end{figure}

\begin{figure}[t]
	\centering
     \includegraphics[width=0.75\textwidth, trim = 0 0 0 0]{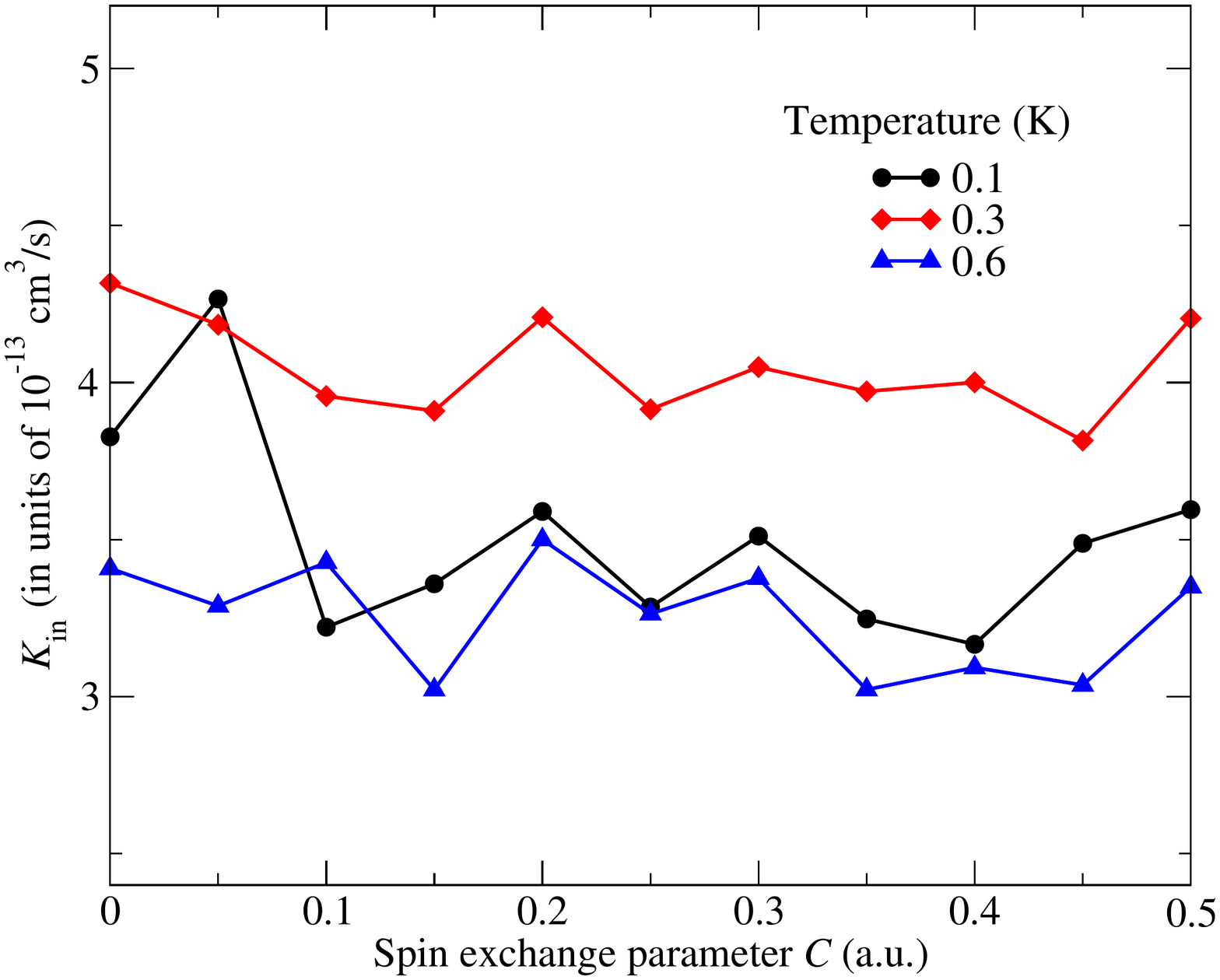}
	\renewcommand{\figurename}{Fig.}
	\caption{Rate constants for dipolar relaxation (\ref{Vaveraging}) calculated for $^{14}$N + $^{14}$N as functions of the SE parameter $C$ for $T=0.6$ K and $C=0$. $\lambda=1$ corresponds to unscaled potential B.}
	\label{fig:NratevsC}
\end{figure}

\begin{figure}[t]
	\centering
     \includegraphics[width=0.75\textwidth, trim = 0 0 0 -50]{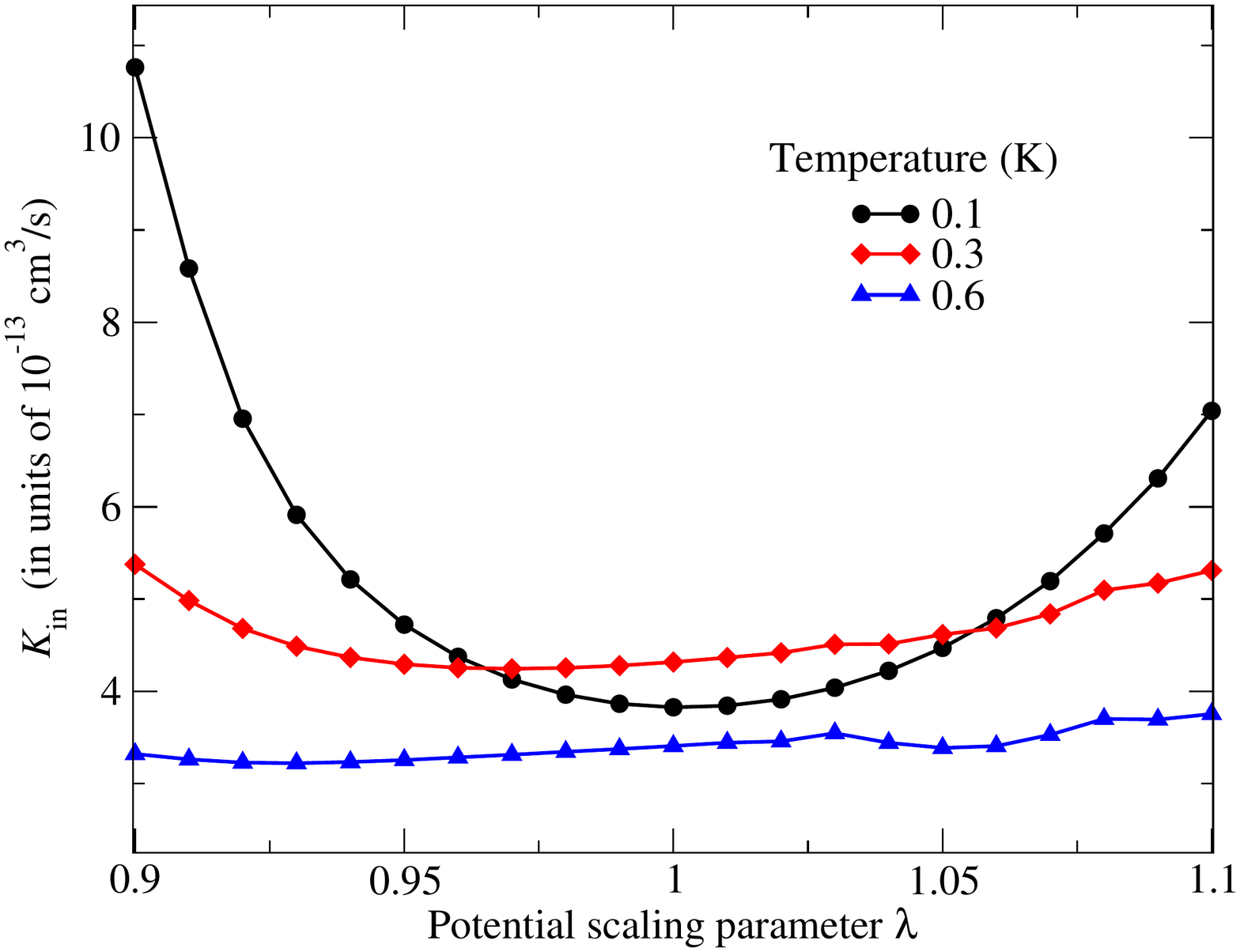}
	\renewcommand{\figurename}{Fig.}
	\caption{Rate constants for dipolar relaxation (\ref{Vaveraging}) calculated for $^{14}$N + $^{14}$N as functions of the potential scaling parameter $\lambda$ for $T=0.6$ K and $C=0$. $\lambda=1$ corresponds to unscaled potential~B.}
	\label{fig:Nratevslambda}
\end{figure}

\begin{figure}[t]
	\centering
     \includegraphics[width=0.70\textwidth, trim = 0 0 0 -70]{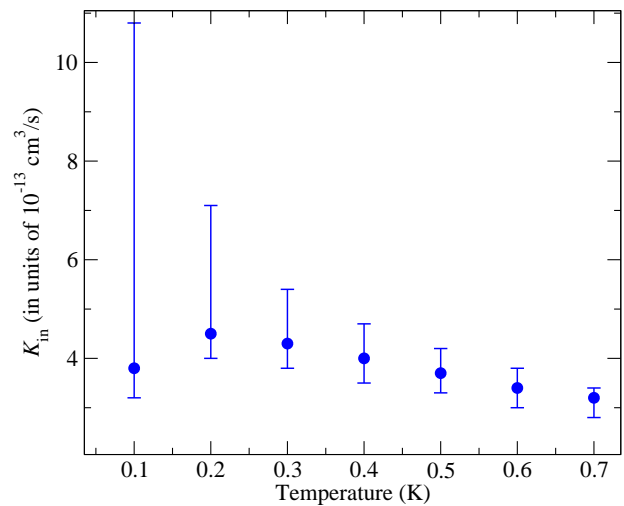}
	\renewcommand{\figurename}{Fig.}
	\caption{Rate constants for dipolar relaxation (\ref{Vaveraging}) calculated for $^{14}$N + $^{14}$N as functions of temperature. The error bars are calculated as explained in the text (see also Table III).}
	\label{fig:NratevsTemp}
\end{figure}

\begin{figure}[t]
	\centering
     \includegraphics[width=0.75\textwidth, trim = 0 0 0 0]{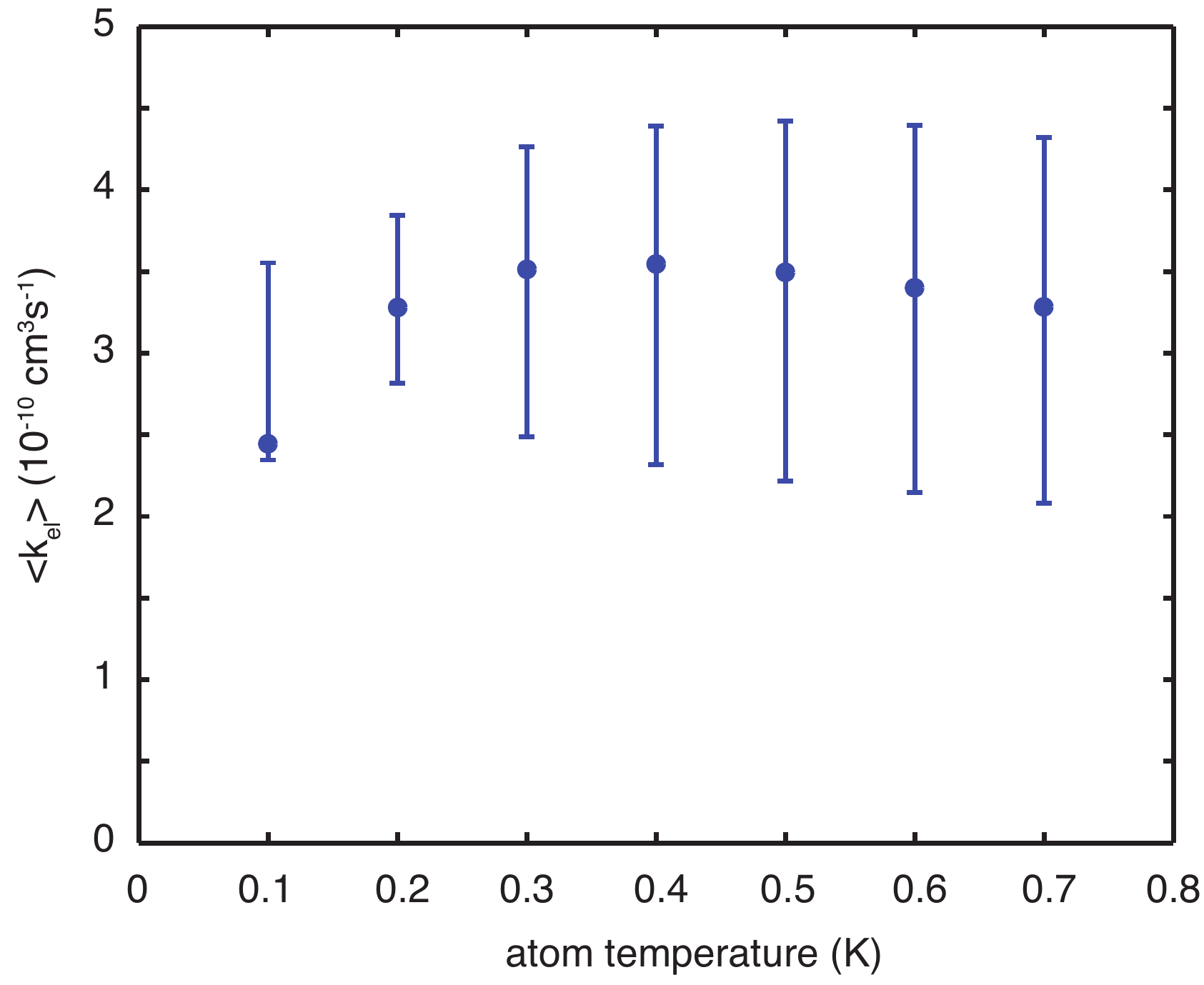}
	\renewcommand{\figurename}{Fig.}
	\caption{Elastic collision rates for $^{14}$N + $^{14}$N averaged over the magnetic field distribution of the trap as functions of temperature.}
	\label{fig:Eratevslambda}
\end{figure}

\begin{figure}[t]
	\centering
     \includegraphics[width=0.7\textwidth, trim = 0 0 0 0]{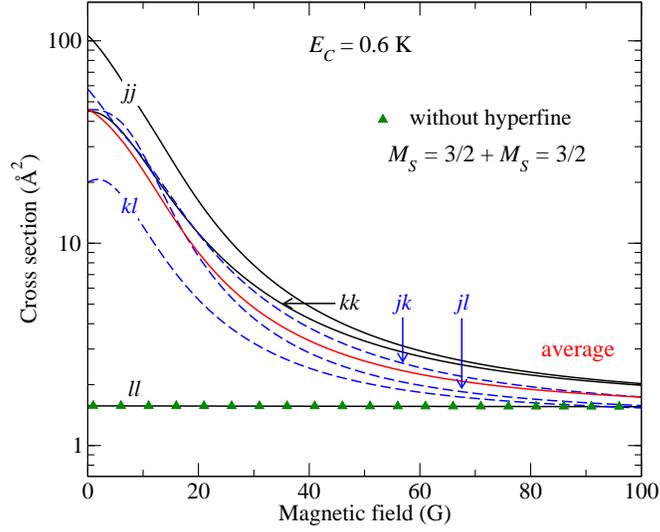}
	\renewcommand{\figurename}{Fig}
	\caption{Cross sections for inelastic relaxation in collisions of $^{14}$N atoms in the low-field-seeking hyperfine states $jj$, $kk$, and $ll$ (full lines) and $jk$, $kl$, $jl$ (dashed lines) calculated for $E_C=0.6$~K and $C=0.5\, E_h$. Also shown are the cross sections calculated with the hyperfine interaction omitted (triangles) for the $|\MSA=3/2\rangle |\MSB=3/2\rangle$ initial channel. The hyperfine levels are labeled as shown in the inset of Fig. \ref{fig:Nlevels}.}
	\label{fig:NcrossvslowB}
\end{figure}

In order to justify the approximation of neglecting the hyperfine structure we made in Sec. IIIA, we performed test calculations of $^{14}$N + $^{14}$N collisions with the hyperfine structure included. The results for $E_C=0.6$ K and $C=0.5\, E_h$ are shown in Fig. \ref{fig:NcrossvslowB}. The inelastic cross sections for N atoms colliding in the uppermost Zeeman state $l$ (Fig. \ref{fig:Nlevels}) are identical to those calculated with the hyperfine structure omitted, as expected for the fully spin-polarized Zeeman states. When the atoms collide in partially polarized Zeeman states $j$ or $k$ at $B<10$ mT, they can exchange spin angular momentum via the SE interaction (\ref{Vsd}). The cross sections for collision-induced SE relaxation are typically much larger than those for dipolar relaxation \cite{Julienne}, so the inelastic cross sections increase by a factor of 50-100 as shown in Fig. \ref{fig:NcrossvslowB}. As $B$ increases, the states $j$, $k$, and $\ell$ converge to the same limit $M_S=3/2$, and the inelastic cross sections decrease monotonically, approaching the same limiting value calculated without taking into account the hyperfine structure (Sec. IIIA). At the temperature and trap depths for the experiments described in Sec. II, N atoms at fields below 10~mT account for less than $10^{-4}$ of the total number of trapped atoms, and therefore do not make a significant contribution to the total trap loss rate.

\newpage


Table I. Parameters and typical values for atomic nitrogen excitation.
\vspace{0.3cm}

\begin{tabular}{l c r l}
\hline\hline
description & symbol & typical value & units \\
\hline 
two-photon cross section \cite{Omidvar_cross_section}  & $\sigma^{(2)}$  & 1.37 & $10^{-36}$~cm$^4$ \\
\hline
excitation pulse energy & $E$ &  $\sim 0.6 $ &mJ\\ 
beam waist & $w_0$  & 120 & $\mu$m\\
effective excitation length & $l_\mathrm{eff}$  & 2 & mm\\
pulse duration (FWHM) & $ \tau_\mathrm{ex} $ & 9.5 & ns\\
resonant line-shape value & $g(0)$ & $(2/\pi) ( 2\pi\times10~$GHz$)^{-1}$ & s\\
2nd-order photon correlation coefficient & $G^{(2)}(0)$ &  2 & \\
\hline
photon collection efficiency &  $\alpha$ & $\sim10^{-4}$ &\\
\hline \hline
\end{tabular}
\label{tab:talif}


\newpage

Table II. Ro-vibrational levels of N$_2(^7\Sigma_u^+$) supported by potentials A and B. The level energies are given in cm$^{-1}$ relative to the N($^4S_{3/2}$) + N($^4S_{3/2}$) dissociation limit in the absence of a magnetic field. The magnetic dipole interaction is not included in the bound-state calculations.
\vspace{0.3cm}

\begin{center}
\begin{tabular}{ccccc}
\hline
\hline
$v,\, \ell$ & \multicolumn{2}{c}{ $^{14}$N$_2$ } & \multicolumn{2}{c}{ $^{15}$N$_2$ } \\
\hline
              &\,\, Potential A \quad & \quad  Potential B \quad & \quad Potential A \quad& \quad Potential B \\
\hline
0,  0  &  -17.38  &  -19.10   & -17.72 \quad  &  -19.47 \\
0,  1  &  -17.08  &  -18.80   & -17.45 \quad  &  -19.19 \\
0,  2  &  -16.50  &  -18.21   & -16.90 \quad  &  -18.63 \\
0,  3  &  -15.62  &  -17.32   & -16.08 \quad  &  -17.80 \\
0,  4  &  -14.46  &  -16.14   & -14.98 \quad  &  -16.69 \\
0,  5  &  -13.01  &  -14.66   & -13.62 \quad  &  -15.30 \\
0,  6  &  -11.28  &  -12.91   & -12.00 \quad  &  -13.66 \\
0,  7  &  -9.28  &  -10.88   & -10.12 \quad  &  -11.74 \\
0,  8  &  -7.02  &  -8.57   & -7.99 \quad  &  -9.58 \\
0,  9  &  -4.51  &  -6.01   & -5.62 \quad  &  -7.16 \\
0,  10  &  -1.76  &  -3.20   & -3.03 \quad  &  -4.51 \\
0,  11  &  --     &  -0.162   & -0.221  \quad  &  -1.64 \\
\hline
1,  0  &  -3.61  &  -4.30   & -4.02 \quad  &  -4.75 \\
1,  1  &  -3.41  &  -4.09   & -3.83 \quad  &  -4.56 \\
1,  2  &  -3.01  &  -3.68   & -3.45 \quad  &  -4.16 \\
1,  3  &  -2.43  &  -3.08   & -2.87 \quad  &  -3.58 \\
1,  4  &  -1.67  &  -2.29   & -2.15 \quad  &  -2.82 \\
1,  5  &  -0.75  &  -1.32   & -1.25 \quad  &  -1.88 \\
1,  6  &    -- &  -0.205   & -0.221 \quad  &  -0.79 \\
\hline
2,  0  &  -0.066  &  -0.125   & -0.129 \quad  &  -0.217 \\
2,  1  &  -0.011  &  -0.055   & -0.062 \quad  &  -0.137 \\
\hline
\hline
\end{tabular}
\end{center}

\vspace{3cm}

Table III. The calculated rate constants $\langle K_\text{in}\rangle$ for $^{14}$N (in units of $10^{-13}$ cm$^3$/s) vs temperature (in~K). The maximum relative uncertainties with respect to the mean value calculated for potential~B, $C=0$ and $\lambda=1$ are presented in the third column. The error bars are based on two sets of calculations using (i) $\lambda = 0.90,\ldots,1.10$ with a grid spacing of 0.01 for $C=0$ and (ii) $C=0,\ldots,0.5$ $E_h$ with a grid spacing of 0.05 $E_h$ for $\lambda=1$. Also indicated are the dominant sources of uncertainty in the calculated rates arising from inaccuracies in the $^7\Sigma_u^+$ interaction potential (IP) and omission of the SE interaction (SE).

\vspace{0.6cm}
\begin{center}
\begin{tabular}{cccc}
\hline
\hline
Temperature  &  \quad  $\langle K_\text{in}\rangle$ \quad & \quad  Uncertainty (\%)  &  Source  \\
\hline
0.1   & \quad  $3.8^{+7}_{-0.6}$   &  184.2   & IP \\
0.2   & \quad  $4.5^{+2.6}_{-0.5}$    & 57.8  & IP \\
0.3   & \quad  $4.3^{+1.1}_{-0.5}$    & 25.6  & IP \\
0.4   &  \quad $4.0^{+0.7}_{-0.5}$    & 17.5 & IP\\
0.5   & \quad  $3.7^{+0.5}_{-0.4}$    & 13.5 & IP \\
0.6   & \quad  $3.4^{+0.4}_{-0.4}$    &  11.8 &  IP and SE \\
0.7   & \quad  $3.2^{+0.2}_{-0.4}$   & 12.5 &  SE \\
\hline
\hline
\end{tabular}
\end{center}

\newpage

\end{document}